\documentclass[conference,10pt]{IEEEtran}
\IEEEoverridecommandlockouts

\usepackage{cite}
\usepackage{amsmath,amssymb,amsfonts}
\usepackage{algorithmic}
\usepackage{graphicx}
\usepackage{textcomp}
\usepackage{soul}
\usepackage[normalem]{ulem}
\usepackage{xcolor}
\usepackage{hyperref}
\usepackage{fancyhdr}
\def\BibTeX{{\rm B\kern-.05em{\sc i\kern-.025em b}\kern-.08em
    T\kern-.1667em\lower.7ex\hbox{E}\kern-.125emX}}

\begin{document}

\newcommand{\zd}[1]{{\color{red}{[LZD: #1]}}}
\newcommand{\rev}[1]{\textcolor{black}{#1}}
\newcommand{\shep}[1]{\textcolor{red}{#1}}
\newcommand{\remove}[1]{\textcolor{red}{\sout{#1}}}


\pagenumbering{arabic}

\title{Photonic Quantum Computing on Spin Memory Architecture with Tree-Encoded Fusion}
\author{
\IEEEauthorblockN{
Xiangyu Ren\IEEEauthorrefmark{2}\IEEEauthorrefmark{1},
Yuexun Huang\IEEEauthorrefmark{3}, 
Zhemin Zhang\IEEEauthorrefmark{4},
Yuchen Zhu\IEEEauthorrefmark{6},\\
Tsung-Yi Ho\IEEEauthorrefmark{4},
Antonio Barbalace\IEEEauthorrefmark{2},
Zhiding Liang\IEEEauthorrefmark{4}\IEEEauthorrefmark{1}
}

\IEEEauthorblockA{
\IEEEauthorrefmark{2}
University of Edinburgh, Edinburgh, Scotland, UK\\
\IEEEauthorrefmark{3}
University of Chicago, Chicago, IL, USA\\
\IEEEauthorrefmark{4}
The Chinese University of Hong Kong, Sha Tin, Hong Kong\\
\IEEEauthorrefmark{6}
Northwestern University, Chicago, IL, USA\\
}
}

\maketitle

\begin{abstract}
Photonic quantum computer (PQC) is a promising quantum computation platform, realizing the measurement-based quantum computation (MBQC) model. 
In MBQC, computation proceeds by preparing a graph state, and this preparation mainly relies on fusion operations.
However, fusion operations on a PQC are prone to two types of errors: fusion \textit{failure} and fusion \textit{erasure}. 
Therefore, the MBQC compiler must be carefully designed to tolerate these errors. 
Previous state-of-the-art MBQC compiler -- OneAdapt, is tailored to all-photonic architectures and primarily address fusion failures. 
However, it neglects fusion erasure errors caused by photon loss, which are more detrimental than fusion failures.

To address the challenge of fusion erasure, we propose a novel MBQC scheme that is based on \textit{quantum spin memory} architecture. 
We design a tree-encoded fusion scheme that effectively suppresses erasure errors. 
Then, we integrate this scheme into our compiler framework, with compilation algorithms that reduce execution overhead of quantum programs. 
We evaluate our framework on a realistic PQC simulator across six typical quantum algorithm benchmarks with various program sizes. 
Our results show that the proposed tree-encoding scheme outperforms other fusion encoding schemes, and that our compilation framework outperforms OneAdapt with exponential improvement. 
Moreover, we demonstrate a small-scale QAOA experiment on real PQC hardware, where it outperforms the latest superconducting hardware.
\end{abstract}


\section{Introduction}
Photonic quantum computer (PQC) is one of the highly potential quantum computing systems that pave the way to quantum supremacy\cite{o2009photonic}. 
Operated at room temperature, the photonic qubits have long decoherence time and great scalability\rev{\cite{rudolph_why_2017}}; also, their natural distributed characteristic makes PQC easy for quantum network integration\rev{\cite{zhan_performance_2023,borregaard_one-way_2020}}.
Various companies have demonstrated their PQC hardware platforms: PsiQuantum introduces their manufacturable platform of PQC with 99.72\% two qubit fusion fidelity\cite{psiquantum_team_manufacturable_2025}, and Quandela provides cloud platform access with a 24-photon modes PQC system~\cite{quandela_cloud}.
Despite the rapid progress of photonic quantum hardware, compilation techniques of photonic quantum computer (PQC) is yet to be fully explored, while there are only a few novel PQC compilers~\cite{zhang_oneperc_2024,zhang_oneadapt_2025} evaluated on a simplified quantum simulation model.

\textbf{MBQC and graph state.} Different from other typical quantum computing hardware (e.g., superconducting or neutral-atoms), photonic quantum computing adapts a measurement-based computation (MBQC) model\cite{briegel2009measurement} rather than the usual gate-based model.
In the paradigm of measurement-based quantum computation, a quantum program is represented as a highly entangled quantum state -- the graph state, and sequential measurements are performed on qubits of the graph state to execute corresponding computations.
Hence, the graph state plays a key role in MBQC, and the main challenge of MBQC compilation on photonic systems is to generate the target graph state robustly and efficiently.

\textbf{Fusion operations for graph states are imperfect.} In a photonic system, photonic qubits are emitted from the photon sources, and they are entangled through \textit{fusion} operations to form a graph state.
Fusion operations are prone to two types of errors: \textit{fusion failure} and \textit{fusion erasure.}
Fusion failures arise from the natural characteristic of photonic qubits.
When the fusion of two qubits fails, they are automatically measured out and have no impact on the rest of their entangled qubits.
In contrast, fusion erasure is caused by photon loss: one of the fusion qubits cannot be detected by the system, thus the outcome of fusion operation remains unknown.
In real PQC hardware, we observe an erasure rate $p_{eras}\approx10\%$. 
This level of erasure error is detrimental to the execution of quantum programs.
Details are explained in Sec.~\ref{subsec:errors}.

\textbf{Prior Works.} A recent state-of-the-art MBQC compiler -- OneAdapt~\cite{zhang_oneadapt_2025} purpose an efficient framework to execute quantum programs on PQC, and it tackles the fusion failure by utilizing a normalization method.
However, it neglects the occurrence of fusion erasure, which is unable to deal with only using normalization. 
Also, normalization method distillates a valid 2D graph state layer from much larger number of photonic qubits, which fails to efficiently utilize photon resources.
Meanwhile, Li et al.~\cite{li_reinforcement_2025} propose the RLGS compiler on emitter-based photonic architecture. 
It benefits from the deterministic nature of emitter-based photon source, and utilize the interactions between emitters to generate entanglements for graph state.
Despite this, emitter-based solution has still not been fully demonstrated in experiment, due to the challenge of coupling between emitters efficiently\cite{hilaire_near-deterministic_2023}.
As a result, the evaluations in \cite{li_reinforcement_2025} are based on assumptions from theoretical analysis~\cite{russo_photonic_2018, gimeno-segovia_deterministic_2019}, so the proposed solution still has a gap from near-term photonic quantum computing systems.


\textbf{Quantum spin memory architecture.} To fill this gap, we introduce our PQC framework to handle the fusion errors in the near-term PQC architecture with a realistic hardware model. 
The PQC architecture that we adapt\cite{hilaire_near-deterministic_2023} utilizes silicon-based \textit{quantum spin memory} as photon source, which could generate a specific structure of resource graph state, namely \textit{caterpillar} state.
The caterpillar state has been demonstrated experimentally in a small scale\cite{huet_deterministic_2025}.
Next, these caterpillar states are concatenated together by fusion operations, forming a large graph state corresponding to the target quantum program.
\ul{Generally, the main problem we are addressing in this work, is finding an efficient solution that generates the target graph state from caterpillar state, while maintaining robustness against fusion failure and erasure.}

Firstly, we propose a \textbf{novel tree-encoded scheme to boost the success rate of fusion operation}, while effectively suppressing fusion failure and fusion erasure.
We leverage the flexibility of caterpillar state, and are inspired by the quantum error correction (QEC) tree code\cite{varnava_loss_2006, albert_tree_2024, bell_optimizing_2023} to design our unique fusion scheme.
Specifically, we eliminate the photonic qubit that undergoes erasure (photon loss) utilizing the \textit{indirect Z-measurement}.
Meanwhile, our dedicated tree structure design provides multiple trials for the fusion operation, offering resistance to fusion failures.
Compared to previous boosted-fusion schemes, e.g. repetition encoding\cite{hilaire_near-deterministic_2023} and repeat-until-success\cite{gliniasty_spin-optical_2024} approach, our tree-encoded scheme achieves significant error-tolerance against high erasure error rates (Fig.~\ref{fig:treecode}(f)). 
Furthermore, the tree-encoding structure can be easily generated from the caterpillar state, proving its fitness for quantum spin memory architecture.
Detailed theoretical analysis and simulation are given in Sec.~\ref{sec:encoding}.

Secondly, we propose \textbf{MemTree} -- a \textbf{compilation framework for scalable and resource-efficient execution} of quantum programs on quantum spin memory architecture.
We implement a divide-and-conquer algorithm to separate the target graph state into multiple caterpillar states while reducing the total number of fusions to minimize execution overhead.
Specifically, we design a hierarchical minimal-cut algorithm for dividing the target graph state.
In the algorithm, we leverage mix-integer-programming (MIP) to obtain the solution, with dedicated constraints to guarantee the divided subgraphs conform to the structure of caterpillar states.
Details of compiler design are given in Sec.~\ref{sec:compiler}.

Thirdly, we build a \textbf{realistic error-aware simulator} to evaluate our framework and compare it with SOTA\cite{zhang_oneadapt_2025}.
Our simulator considers the impact of fusion failure and fusion erasure based on the hardware noise model, setting the configurations from experimental works\cite{huet_deterministic_2025, psiquantum_team_manufacturable_2025, maring_versatile_2024, bartolucci_fusion-based_2023, thomas_fusion_2024}.
We choose two dominant overheads of photonic MBQC as metrics: (1) Total execution time of the quantum program. (2) Number of photon sources required to perform the computation.
In addition, our simulation considers the errors in preparing tree-encoded logical qubits, ensuring a realistic evaluation of the fault-tolerance scheme.
Details are given in Sec.~\ref{sec:methodology}.

Lastly, we perform the \textbf{evaluation on a comprehensive set of benchmarks}, which includes 6 quantum algorithms, each with varying sizes (36-100 qubits).
We evaluate our tree-encoded fusion scheme against previous boosted fusion scheme -- repetition-encoded fusion\cite{hilaire_near-deterministic_2023} and repeat-until-success fusion\cite{lim_repeat-until-success_2005}, where our scheme reduces the program execution time by a factor of $1.9\times10^{-3}$ and $1.7\times10^{-2}$, respectively.
We compare our PQC framework with OneAdapt, and on average, we achieve reduction rates of $1.5\times10^{-2}$ in execution time, $0.18\times$ in photon resources, $0.14\times$ in compilation runtime, and a $3.64\times$ improvement in fidelity.
Compared with RLGS\cite{li_reinforcement_2025}, we achieve an improvement of $1.42\times$ in fidelity.

Our contributions are listed as follows:

1. We consider a more realistic error mechanism in photonic quantum computing and propose a tree-encoded fusion scheme to protect against fusion failure and erasure, while the latter error has been overlooked by previous works\cite{10.1145/3579371.3589047,zhang_oneperc_2024,zhang_oneadapt_2025}.

2. We tailor the tree-encoded fusion scheme to a novel PQC architecture -- the spin memory architecture, and design a compilation framework that reduces program execution time while minimizing photon resource overhead.

3. We implement a realistic simulator of spin memory architecture, which is based on configurations of a successfully demonstrated hardware platform\cite{huet_deterministic_2025}. We compare our work with other fusion schemes\cite{hilaire2023near,gliniasty_spin-optical_2024} and SOTA PQC compilers\cite{zhang_oneadapt_2025}, while the results show significant improvements in execution time and resource overhead.

4. To the best of our knowledge, prior compiler works for photonic MBQC have been evaluated primarily in simulation, whereas our work includes real-hardware validation through a quantum algorithm demonstration, and shows improvement compared to prevalent superconducting hardware.

\section{Background}
\subsection{MBQC Background}
\textbf{The Graph state} is a special type of multipartite entangled state~\cite{hein2006entanglement}, with its intrinsic structure determined by a graph $G=(V,E)$, where $V$ denotes the set of vertices and $E$ denotes the set of edges. Then, each vertex $v\in V$ is associated with a qubit. The formal definition of a graph state $|G\rangle$ is given by
\begin{equation}
\label{eq:graphState}
    \vspace{-1mm}
    |G\rangle=\prod_{(i,j)\in E} CZ_{(i,j)}|+\rangle^{\otimes V},
    \vspace{-1mm}
\end{equation}
where $|+\rangle^{\otimes V}$ is the tensor product state with all $|V|$ qubits initialized in the $X$ eigen-state $|+\rangle$, and $CZ_{(i,j)}$ represents the CZ gate applied to qubits(vertices) $i$ and $j$ connected by edge $(i,j)$ in $G$. 
The graph state is shown to be universal for \textbf{MBQC} \cite{raussendorf2001one}, in the sense that only single qubit measurements are required for any computation once the graph state is generated. 
We refer to the background sections of \cite{zhang_oneperc_2024, zhang_oneadapt_2025} for more details about graph states and MBQC.

\begin{figure}[t]
    \centering
    \includegraphics[width=0.48\textwidth]{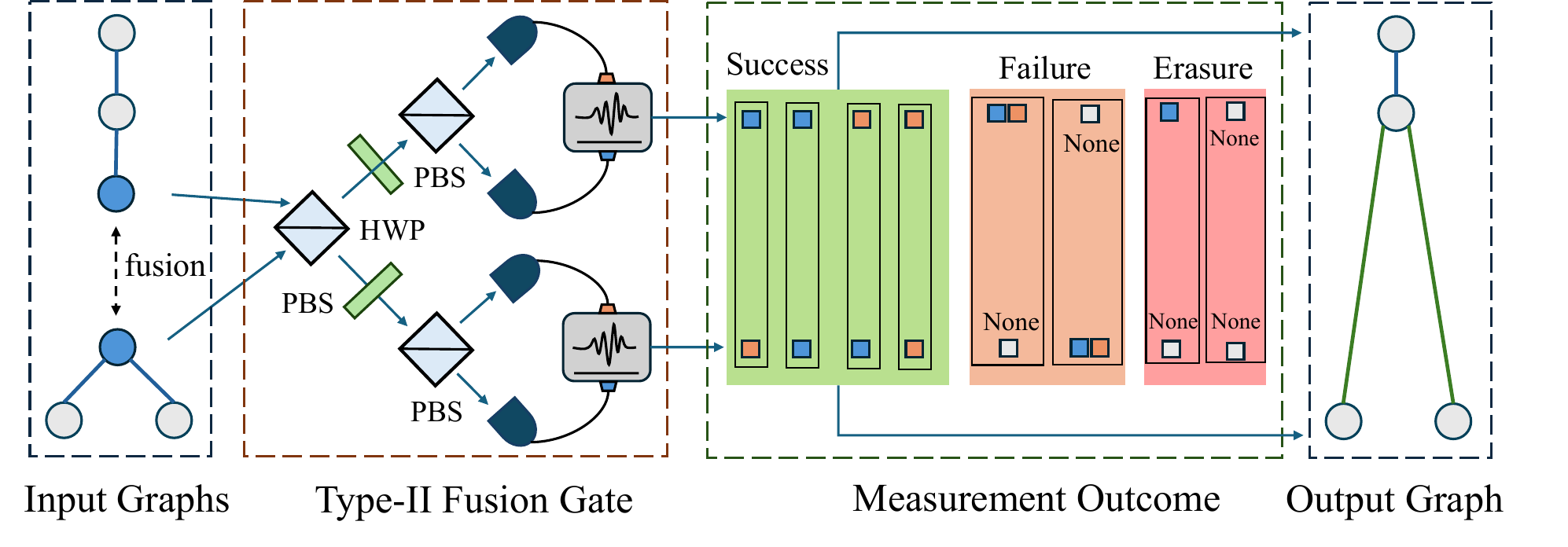}
    \caption{Type-II fusion operation. Two qubits (each from one input graph) are performed fusion operation, combined into a larger output graph. The fusion is success or not depends on the measurement outcome of these two qubits: if they are captured in different sides of detectors, the fusion succeeds; if captured at the same side, the fusion fails; if one of the qubits not captured, it leads to fusion erasure.}
    \label{fig:type2fusion}
    \vspace{-2mm}
\end{figure}

\textbf{Fusion} is arguably the most important operation in graph state generation, as it allows us to combine smaller graph states to form a desired larger target graph state, thus enabling resource efficient parallel generation \cite{browne2005resource,hoyer2006resources}. 
In the \textit{Type-I fusion}, the two vertices entering the fusion gate from the two smaller graph states are merged into one, inheriting the edges from both and resulting in a larger graph state. 
Whereas the result for \textit{Type-II fusion} is a bit more complicated: both input vertices are removed from the graph, while the neighbors of one vertex are connected to (disconnected from) the neighbors of the other vertex if they were previously disconnected (connected). 
Both types of fusion can be implemented probabilistically via linear optics using half-wave plates (HWP) and polarizing beam splitters (PBS). 
And we focus on Type-II fusion in this work, as its photon loss can be heralded \cite{li2015resource}, for which an intuitive example\footnote{The implementation of type fusion varies across different literature depending on whether HWP are inserted before the first PBS \cite{browne2005resource,lee2023graph,hilaire2023near}, resulting in different local unitary corrections required upon success and different effective measurements upon failure. 
We adapt the scheme in \cite{hilaire_near-deterministic_2023}.} is provided in Fig.~\ref{fig:type2fusion}.

\subsection{PQC Hardware Architectures}
\begin{figure}[t]
    \centering
    \includegraphics[width=0.47\textwidth]{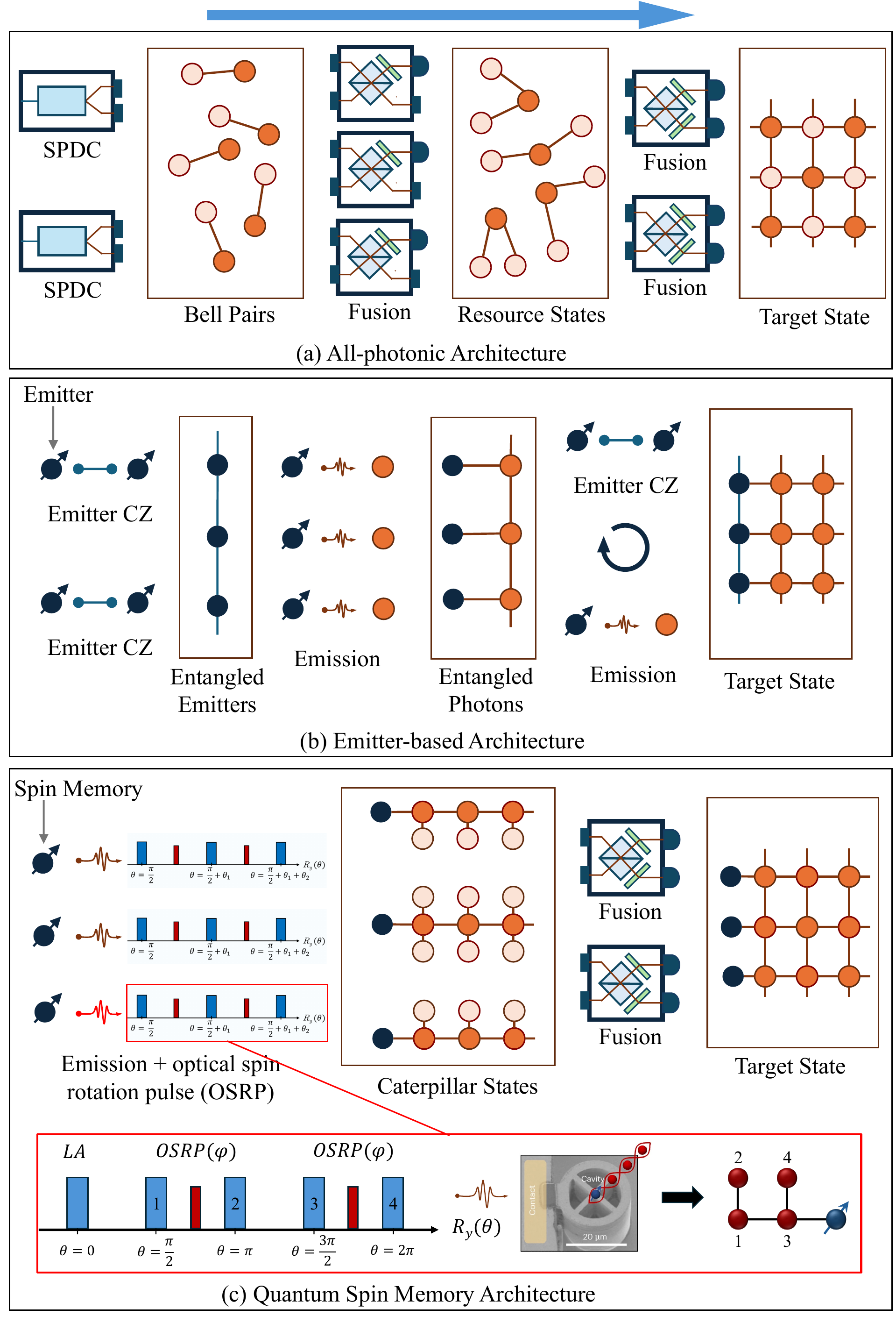}
    \caption{The comparison among different PQC architecture and their corresponding graph state generation schemes. \rev{The excitation pulses for generating a caterpillar state are demonstrated in the red box of (c). Specifically, longitudinal-acoustic excitation $LA(\frac{\pi}{2})$ and optical spin rotation pulses ($OSRP(\varphi),\varphi=\pi$) are applied to QD-cavity~\cite{huet2025deterministic} in dedicated sequence, emitting the target caterpillar graph shown on the right.}}
    \label{fig:ArchComp}
    \vspace{-2mm}
\end{figure}

The key and most difficult part of generating the graph state is constructing the required CZ connections (edges) between the qubits (vertices). There are mainly three fundamental hardware architectures for generating the resource graph states for PQC, as illustrated in Fig.~\ref{fig:ArchComp}(a)-(c). 

In the \textbf{all-photonic} architecture, the required equipment consists of linear optical elements and the conventional spontaneous parametric down-conversion (SPDC) source for photonic bell pair generation \cite{browne2005resource,kwiat1995new,zhuang2025ultrabright}. 
The generated photonic bell pairs can be merged together to form a larger graph using the aforementioned fusion operation. 
However, such a fusion process is probabilistic, with a $50\%$ chance at best (which can be enhanced to $75\%$ by additional optical hardware\cite{grice2011arbitrarily,ewert20143}). 
As illustrated in Fig.\ref{fig:ArchComp}(a), a graph state with any underlying topology can be generated by repeating the above fusion process with sufficient resource bell pairs \cite{lee2023graph}, which is the scheme explored in OneAdapt\cite{zhang_oneadapt_2025}.

In contrast, the \textbf{emitter-based} architecture theoretically promises a deterministic generation of graph states through the use of interacting quantum emitters\cite{lindner2009proposal,schon2005sequential,economou2010optically,gimeno2019deterministic}. The proposal arises from two basic mechanisms shown in Fig.~\ref{fig:ArchComp}(b): (1) The emitted photon from a quantum emitter is entangled with this emitter, resulting in an effective CNOT gate between the emitter qubit and the photon qubit. (2) The emitters themselves can be entangled with each other by implementing a CZ gate, which is theoretically analyzed in \cite{russo_photonic_2018}, but has not yet been experimentally demonstrated. These mechanisms result in a series of basic generation rules, based on which an arbitrary graph state can be generated\cite{kaur2024resource,ren2025scalable,li2022photonic}. RLGS\cite{li_reinforcement_2025} is the SOTA compiler framework based on this architecture.

The \textbf{quantum spin memory} is a PQC architecture\cite{gliniasty_spin-optical_2024,chan_tailoring_2025,huet2025deterministic,pettersson_deterministic_2025} based on semiconductor quantum dot (QD) emitters.
The preparation of a caterpillar state is illustrated in the red box of Fig.~\ref{fig:ArchComp}(c).
When applying longitudinal-acoustic (LA) excitation pulses on the QD-cavity iteratively, it can emit linearly entangled photons as graph states.
Additionally, we can interleave optical spin rotation pulses (OSRP) into the excitation pulses, leading to the emission of a special graph state structure -- the \textit{Caterpillar} state\cite{pettersson_deterministic_2025}.
As shown in Fig.~\ref{fig:ArchComp}(c), \textit{Caterpillar} state has a branched-chain structure, with a chain of linearly entangled qubits as the \textit{main path}, and extra \textit{leaf qubits} each directly connected to one qubit of the \textit{main path}.
We refer to \cite{pettersson_deterministic_2025} for its complete definition and \cite{huet2025deterministic} for its physical preparation process.
Next, the caterpillar states are concatenated into the target graph states through a set of fusion operations using linear optical hardware similar to an all-photonic architecture.
While spin memory architecture is also prone to fusion errors, the flexibility within the caterpillar structure enables us to design and integrate error-tolerant encoded graph states (details in Sec.~\ref{sec:encoding}).

\begin{table*}[h]
\centering
\begin{tabular}{|c|c|l|c|l|}
\hline
Architecture &
  Photonic Hardware &
  \multicolumn{1}{c|}{Error Types} &
  Existing Problem &
  \multicolumn{1}{c|}{Prior Compiler Framework} \\ \hline
all-photonic &
  linear optics &
  \begin{tabular}[c]{@{}l@{}}\textbullet~ fusion failure \\ \textbullet~ fusion erasure\end{tabular} &
  \begin{tabular}[c]{@{}c@{}}\textbullet~ Low utilization rate of photons\\ \textbullet~ Fusion erasure error unsolved\end{tabular} &
  \begin{tabular}[c]{@{}l@{}}\textbullet~ OneAdapt\cite{zhang_oneadapt_2025}, \\ \textbullet~ FCM\cite{mo2024fcm}\end{tabular} \\ \hline
emitter-based &
  quantum emitter &
  \begin{tabular}[c]{@{}l@{}}\textbullet~ emitter decohenrence\\ \textbullet~ emitter-CZ infidelity\end{tabular} &
  \begin{tabular}[c]{@{}c@{}}Bottleneck of experimentally \\ demonstrating the emitter-CZ\end{tabular} &
  \begin{tabular}[c]{@{}l@{}}\textbullet~ RLGS\cite{li_reinforcement_2025}, \\ \textbullet~ GSDiv\cite{ren2025gsdiv}\end{tabular} \\ \hline
spin memory &
  \begin{tabular}[c]{@{}c@{}}quantum spin memory +\\  linear optical hardware\end{tabular} &
  \begin{tabular}[c]{@{}l@{}}\textbullet~ fusion failure \\ \textbullet~ fusion erasure\end{tabular} &
  Fusion erasure error unsolved &
  None \\ \hline
\end{tabular}
\vspace{2mm}
\caption{Comparison between photonic quantum computing architectures.}
\label{tab:arch_comp}
\vspace{-5mm}
\end{table*}

\section{Motivation}
\subsection{Errors in Fusion Operation}\label{subsec:errors}
In type-II fusion, the two dominant error sources are \textit{fusion failure} and \textit{fusion erasure}. 
Although both arise from the same imperfect fusion primitive, they differ in a key way: fusion failure leads to a \emph{known} graph transformation, whereas fusion erasure leads to an \emph{unknown} graph outcome.

As shown in Fig.~\ref{fig:type2fusion}, \textbf{fusion failure} is heralded when two fusion qubits are captured in the same detector, indicating that the desired entanglement is not created. 
In this case, the failed qubits are effectively measured in the Z basis and disconnected from the graph. 
Therefore, although the fusion attempt is unsuccessful, the resulting graph structure remains known to the compiler. 
This is the failure model considered in previous compilers such as OneAdapt~\cite{zhang_oneadapt_2025} and OnePerc~\cite{zhang_oneperc_2024}.

In contrast, \textbf{fusion erasure} is triggered by photon loss during fusion, where one fusion qubit cannot be captured by the detector, as shown in Fig.~\ref{fig:type2fusion}. 
The erased qubit is no longer accessible for computation, and its effect cannot be removed by a direct Z measurement. 
More importantly, the output graph state of the fusion becomes uncertain, since it is unknown whether the entanglement has been established or not. 
Such uncertainty is especially harmful to MBQC, because later measurements rely on the exact graph-state structure; therefore, the corrupted fusion output must be discarded unless additional protection is applied.

\begin{figure}[h]
    \vspace{-3mm}
    \centering
    {\includegraphics[width=0.47\textwidth]{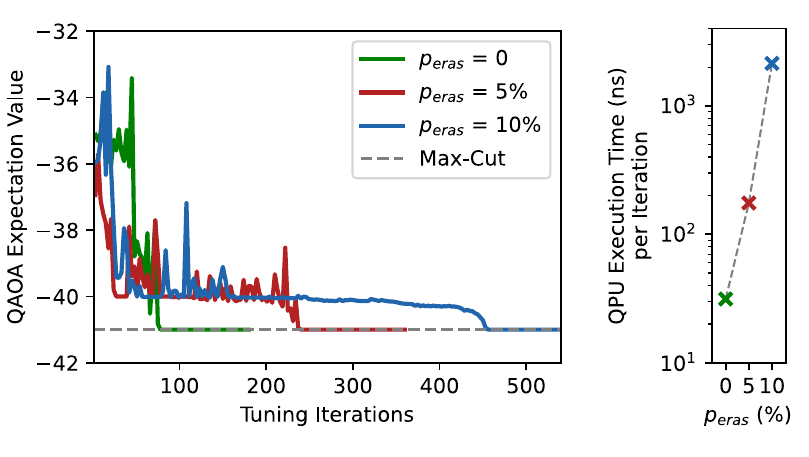}}
    \rev{\caption{Optimizing a Max-Cut problem using 6-qubit QAOA program on PQC simulator\cite{heurtel2023perceval}. We use the RUS boosted fusion method ($m=6$), and simulate the fusion erasure at 0, 5\% and 10\% respectively, while fixing the fusion failure at 25\%. Left: Optimization of QAOA expectation value. Right: Quantum circuit execution time per tuning iteration.}
    \label{fig:motivation}}
    \vspace{-4mm}
\end{figure}

These two errors are also closely related in boosted-fusion design. 
A common way to suppress fusion failure is to introduce more fusion attempts, but each extra attempt also exposes more qubits to photon loss and thus increases the chance of fusion erasure. 
As a result, improving tolerance to fusion failure alone is insufficient under realistic photon-loss conditions.
To demonstrate the impact of erasure, Fig.~\ref{fig:motivation} shows a simulation of a Max-Cut QAOA program under different erasure rates using the previous best fusion scheme. 
The results show that erasure undermines quantum programs in two aspects: 
(i) it increases the \ul{number of fusion attempts}, leading to exponentially \ul{longer execution time}; 
(ii) the longer execution time accumulates \ul{higher decoherence and CZ errors}, resulting in lower-quality outputs and larger program-level overhead, such as more tuning iterations in QAOA.

\subsection{Problems in Previous SOTA Compilers}
OneAdapt\cite{zhang_oneadapt_2025} and RLGS\cite{li_reinforcement_2025} are the SOTA compilers for all-photonic and emitter-based architectures, respectively.
Though carefully designed, there are still several gaps from implementing the realistic error-tolerant MBQC, and we conclude their existing problems in Table.~\ref{tab:arch_comp}.

\textbf{OneAdapt} iteratively generates resource state layers (RSL) and normalizes them into effective 2D layers of lattice graph states to create the target graph state of a quantum program.
This strategy resolves the fusion failure problem, but it overlooks the fusion erasure, which induces errors in the 2D graph state layers.
Assuming an $1\%$ erasure rate, generating the required $84\times84$ RSL will demand $>10^5$ fusion operations, leading to an extremely low probability of not experiencing erasure in the whole RSL.
Furthermore, the normalization method results in a low utilization rate of photons. For example, OneAdapt normalizes only a $4\times4$ 2D layer from the $84\times84$ qubits RSL\cite{zhang_oneadapt_2025}.

The problem of \textbf{RLGS} primarily lies in the bottleneck of emitter-based architecture hardware. 
Up to now, only the generation of linear graph states with few qubits from a single quantum emitter has been experimentally demonstrated \cite{schwartz2016deterministic,cogan_deterministic_2023}. 
The hardware bottleneck arises from the inability to demonstrate high-quality CZ interactions between two emitters\cite{hilaire2023near}, which is essential for generating MBQC graph states. 

\subsection{Potentials in Spin Memory Architecture}
We list our insights on addressing the above challenges in the MBQC compiler, leveraging the spin memory architecture:

(1) The caterpillar state structure offers the chance to resist fusion failure by embedding specific graph state patterns, known as the boosted fusion scheme\cite{hilaire_near-deterministic_2023}. 
Since the  above scheme failed to deal with fusion erasure, in this work, we explore the graph state pattern that tolerates both failure and erasure, while tailoring the pattern to the caterpillar state (Sec.~\ref{sec:encoding}).

(2) Equipped with the error-tolerant graph state pattern, we can improve the fusion success rate. 
As a result, we have no need for excessive photon sources and apply normalization like OneAdapt.
In contrast, we arrange the generation of caterpillar states from photon sources to be \textit{program-agnostic}, while the structure of each caterpillar state is on demand and determined by the target graph state.
Hence, we can improve the utilization rate of photon sources, with details in Sec.~\ref{sec:compiler}.

(3) Our compiler considers the hardware settings from real experiments, gaining more robustness and being more achievable in near-term PQC.
Compared to the emitter-based architecture, which still has unsolved hardware barriers, the spin memory architecture is accessible on cloud platforms\cite{heurtel2023perceval, quandela_cloud}.

\begin{figure*}[t]
    \centering
    \includegraphics[width=0.97\textwidth]{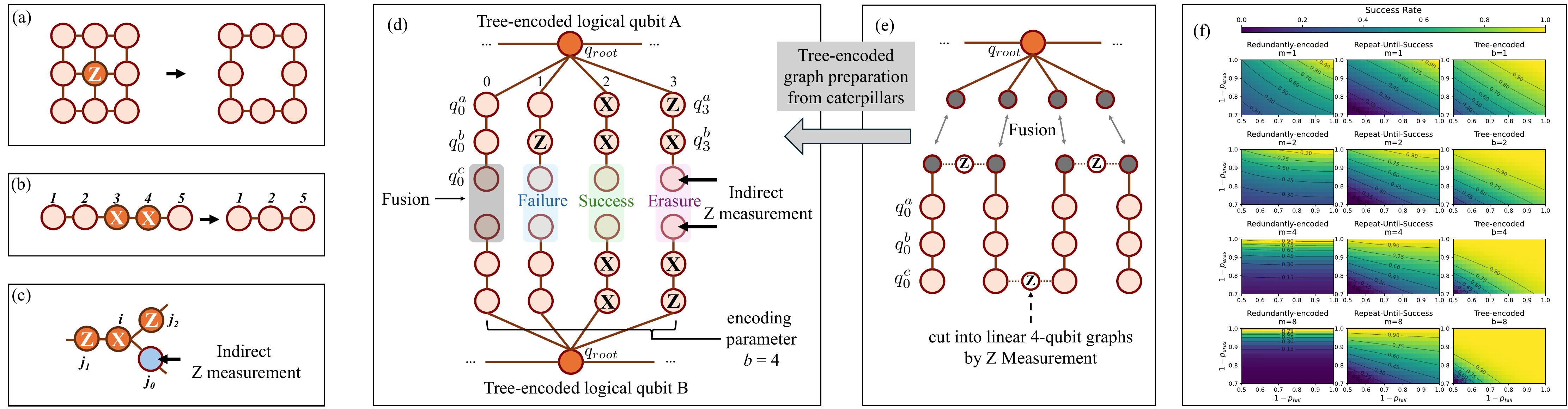}
    \caption{(a)-(c) Graph state measurement patterns that establish loss-tolerance.
    (d) Tree-encoded fusion scheme. (e) Preparing tree-encoded logical qubit from caterpillar states. (f) Simulation of the fusion schemes. We compare these schemes with varying encoding parameters ($m,b=1,2,4,8$), by performing $10^3$ fusion trials per data point to measure success rates.}
    \label{fig:treecode}
    \vspace{-2mm}
\end{figure*}

\section{A Robust Tree-Encoding Fusion Scheme}
\label{sec:encoding}
\subsection{Previous Boosted-Fusion schemes}
First, we briefly describe prior error-tolerant fusion schemes on spin memory architecture and analyze their problems.

\textbf{Redundantly-encoded fusion.}
Hilaire et al.\cite{hilaire_near-deterministic_2023} introduce a fusion failure-tolerant protocol to improve the success rate.
Utilizing the feature of the caterpillar state, the redundantly-encoded scheme generates a logical linear graph state. 
Each node of this linear graph is a logical qubit encoded by $m$ leaf qubits.
For the fusion of two logical qubits from different graph states, a fusion operation is applied to each pair of leaf qubits from these two logical qubits.
As a result, $m$ attempts are performed between the two logical qubits, while any single successful attempt leads to fusion success. 
The redundant fusion operations reduce the overall rate of fusion failure, however, they lead to more exposure to fusion erasure.
Assuming $p_{fail}$ and $p_{eras}$ as the probabilities of fusion failure and erasure, respectively, an $m$-qubit redundantly-encoded fusion has the following logical error rates:
$$ P_{fail} = {p_{fail}}^m,~
P_{eras} = 1 - {(1 - p_{eras})}^{2m}.$$
$P_{eras}$ has a $2m$ exponential because any one of the $m$ physical qubits in each logical qubit is exposed to erasure error independently\cite{gliniasty_spin-optical_2024,chan_tailoring_2025}.
Here we can observe that when the redundant encoding parameter $m$ grows, the logical failure rate $P_{fail}$ reduces, but the logical erasure rate $P_{eras}$ increases.

\textbf{Repeat-until-success fusion.}
Lim et al.\cite{lim_repeat-until-success_2005} propose a repeat-until-success (RUS) method to improve the fusion operation\cite{gliniasty_spin-optical_2024,thomas_fusion_2024}.
In this RUS scheme, ancillary photons are used to apply fusion operations between two photon sources, thus creating entanglement between the caterpillar states produced by the two photon sources.
The RUS scheme has a similar idea to the redundantly-encoded scheme, but it terminates once the fusion operations succeed.
Its logical error rates are:
$$ P_{fail} = {p_{fail}}^m, ~P_{eras} = \sum_{i=0}^{m-1} {p_{fail}}^i \cdot 2p_{eras}$$
Although the RUS scheme achieves slightly better performance than the redundantly-encoded scheme\cite{hilaire_near-deterministic_2023}, it consumes more ancilla qubit resources and takes a longer time.
Moreover, it has the same problem of intolerance to erasure errors.

\subsection{Preliminary: Loss-Tolerant Graph State Patterns} 
Before introducing our tree-encoded fusion scheme, we explain the underlying principles.
Our insight into tree-encoded fusion derives from the QEC tree graph-state code\cite{varnava_loss_2006, albert_tree_2024,bell_optimizing_2023}.
It reveals several properties of graph state measurement:
Fig.~\ref{fig:treecode}(a) shows the \ul{direct Z measurement} rule:
A Z-basis measurement removes the target qubit from its graph state and breaks all the entanglement between the target qubit and all other qubits in the graph state.
Fig.~\ref{fig:treecode}(b) shows the \ul{pair of X measurement} rule:
Two adjacent X-basis measurements on a linear cluster remove the qubits and form direct bonds between their neighbors.
Most importantly, Fig.~\ref{fig:treecode}(c) shows the \ul{indirect Z measurement} rule:
we select a neighboring qubit $i$ of the target qubit $j_0$ and perform an X measurement; then we perform Z measurements on all other qubits $j_1,j_2$ connected to this neighboring qubit $i$.
These measurements deterministically reveal what the Z measurement outcome would have been on $j_0$, based on the underlying stabilizer operator $X_i\prod_{j\in E(i)}Z_j$.
Hence, if the target qubit $j_0$ that we want to measure undergoes photon loss, we can still measure it indirectly with this measurement pattern.
We recommend \cite{varnava_loss_2006} for complete details of the QEC tree code protocol.
 


\subsection{Design of Tree-Encoded Fusion.}
Inspired by both the redundantly-encoded fusion and the QEC tree code, we introduce our tree-encoded fusion scheme, illustrated in Fig.~\ref{fig:treecode}(d). 
Two logical qubits $A$ and $B$ involved in a fusion operation are encoded by a tree-structure: the root qubit $q_{root}$ is connected to $b$ branches, while each branch contains a linear graph of 3 qubits -- $\{q_i^a, q_i^b, q_i^c\}$.
The leaf qubit $q_i^c$ is assigned for fusion measurement, while the other two qubits $q_i^a,q_i^b$ are ancillary qubits for indirect measurement in case of fusion erasure.
The specific operations are listed below for different outcomes of fusion:
\begin{enumerate}
    \item Fusion success: perform the pair of X measurements on $q_i^a$ and $q_i^b$, so that the successful fusion entanglement is directly connected to $q_{root}$.
    \item Fusion failure: upon fusion failure, the $q_i^c$ is measured out, while $q_i^a,q_i^b$ remains in the tree. Then we apply the Z measurement on $q_i^b$ to remove it and leave $q_i^a$ for backup usage, as explained in (4).
    \item Fusion erasure: if $q_i^c$ undergoes erasure, we apply an X measurement on $q_i^b$ and a Z measurement on $q_i^a$, leading to an indirect Z measurement on $q_i^c$. Such operations effectively eliminate $q_i^c$ without impacting the rest of the qubits, hence forming erasure-tolerance.
    \item In the extreme case that all branches of the tree end up with fusion failure or erasure, the backup $q_i^a$ in (2) can still be used for a fusion attempt.
\end{enumerate}
Owing to the protection of fusion erasure, the logical qubits $A$ and $B$ retain an entanglement when one of the branches $i$ succeeds in fusion.
It should be noted that our design of the tree-encoded scheme is a trade-off between the fusion success rate and photon resource consumption.



\subsection{Analysis of Tree-Encoded Fusion.}
Here, we analyze the fusion success rate of redundantly-encoded, RUS, and tree-encoded schemes under certain fusion failure and erasure probabilities (denoted by $p_{fail}$ and $p_{eras}$).
With $m$ or $b$ as the encoding parameter, we calculate their theoretical success rates as:
$$ S_{redun} = (1 - {p_{fail}}^m) \cdot (1-p_{eras})^{2m}$$
$$ S_{rus} = 1 - \sum_{i=0}^{m-1} {p_{fail}}^i \cdot 2p_{eras} - {p_{fail}}^m$$
$$ S_{tree} = 1 - {(1-{(1-p_{eras})}^2 + p_{fail})}^b$$
Here in $S_{tree}$, the probability of fusion erasure in each branch is calculated as $P_{eras} = 1-{(1-p_{eras})}^2$, since the probability of no erasure on each side is $1-p_{eras}$.

In Fig.~\ref{fig:treecode}(f), we simulate the process of these fusion schemes, following the error probabilities and counting the average success rate.
We show the success rate of the logical fusion operation, with $1-p_{fail}$ and $1-p_{eras}$ as the X and Y axes.
It can be observed that our tree-encoded scheme outperforms the redundant-encoded scheme and RUS when the erasure rate $p_{eras}$ scales up.

\subsection{Tailoring Tree-Encoded Fusion to Spin Memory}
The tree-encoded scheme can be generated from caterpillar states on a quantum spin memory architecture.
In Fig.~\ref{fig:treecode}(e), we demonstrate the process of ensembling the tree-encoded logical qubit from caterpillar states.
First, we generate the caterpillar state that is used for the combination into the target state. 
Each logical qubit involved in later fusion is composed of $q_{root}$ on \textit{the main path} and $b$ leaf qubits (in gray color) connected to $q_{root}$.
Meanwhile, we generate $b$ 4-qubit linear graph states,
which can be separated from a long linear graph by using Z measurement.
Then, we apply fusion to concatenate these linear graph states to the leaf qubits appended to $q_{root}$, and form the tree-encoded structure we need for fusion.

\rev{
Since the preparation procedure for tree-encoded logical qubit is not protected by the scheme itself, preparations of these tree branches are exposed to fusion errors. 
To ensure a steady and robust preparation of a logical qubit with $b$-branching, we introduce a \textit{preparation parameter} $b_{prep}$ complying with $b_{prep}>b$.
During the preparation, we perform $b_{prep}$ attempts of branch preparation: (i) If fusion failure happens to a branch, it will be measured out automatically. (ii) If fusion erasure happens to a branch, we solve it by indirect measurement with Z-measurement on $\{q_i^b,q_i^c\}$.
All the branch preparation are performed simultaneously in one timestep, if less than $b$ branches are prepared successfully among $b_{prep}$ attempts, it will be retried in the next timestep.
In the next subsection we discuss an appropriate selection of $b$ and $b_{prep}$ under near-term PQC hardware limitation.
}

\begin{figure}[t]
    \centering
    {\includegraphics[width=0.45\textwidth]{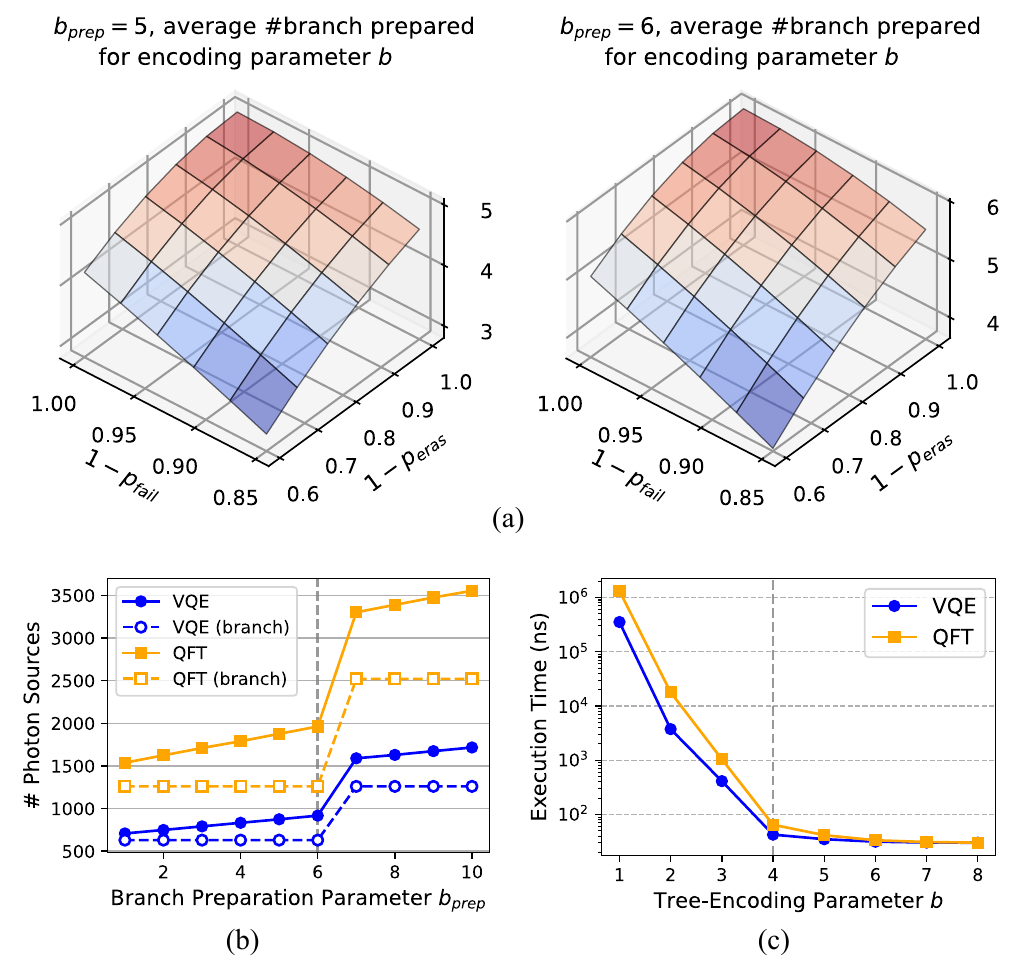}}
    \vspace{-2mm}
    \caption{\rev{(a) Average number of tree branches that successfully prepared for logical qubit encoding parameter $b$, when the preparation parameter $b_{prep}=5$ and $b_{prep}=6$ (by simulation). (b) Photon resource breakdown analysis for parameter $b_{prep}$, when given the maximum length of caterpillar is 30-qubit. Dashed lines represent the \#photon sources used for branch preparation. (c) Execution time analysis for the tree-encoding parameter $b$, under a noise model that $p_{fail}=2\%$ and $p_{eras}=25\%$.}}
    \label{fig:fprep}
    \vspace{-3mm}
\end{figure}

\rev{
\subsection{Tree-encoding Parameter}
The selection of $b$ and $b_{prep}$ should ensure less preparation timestep and photon sources.
In Fig.~\ref{fig:fprep}(b), we count the \#photon sources required for varying $b_{prep}$.
Given a 30-qubit limitation of caterpillar graph, \#photon sources keeps steady and grows sharply when $b_{prep}>6$.
In Fig.~\ref{fig:fprep}(c), we evaluate the average execution time of MemTree on 36-qubit VQE and QFT programs. The results show that with an increasing value of $b$, execution time decreases on an exponential scale until $b=4$, followed by convergence afterward.
}

Next, we study the relationship between $b_{prep}$ and $b$ in Fig.~\ref{fig:fprep}(a).
Under the realistic fusion error rates of near-term PQC, when $b_{prep}=6$ the preparation can ensure $>4$ branches prepared averagely.
Hence we select $b=4$ and $b_{prep}=6$ for a tradeoff between performance (fusion success rate) and number of photon sources.
Furthermore, in real hardware experiment (Sec.~\ref{subsec:realexp}), we obtain a 83.3\% preparation success rate within one timestep and 97.1\% success rate within in two timesteps.
In the future, these parameters can be adjusted based on maximal capability of caterpillar generation.



\begin{figure*}
    \centering
    \includegraphics[width=0.95\textwidth]{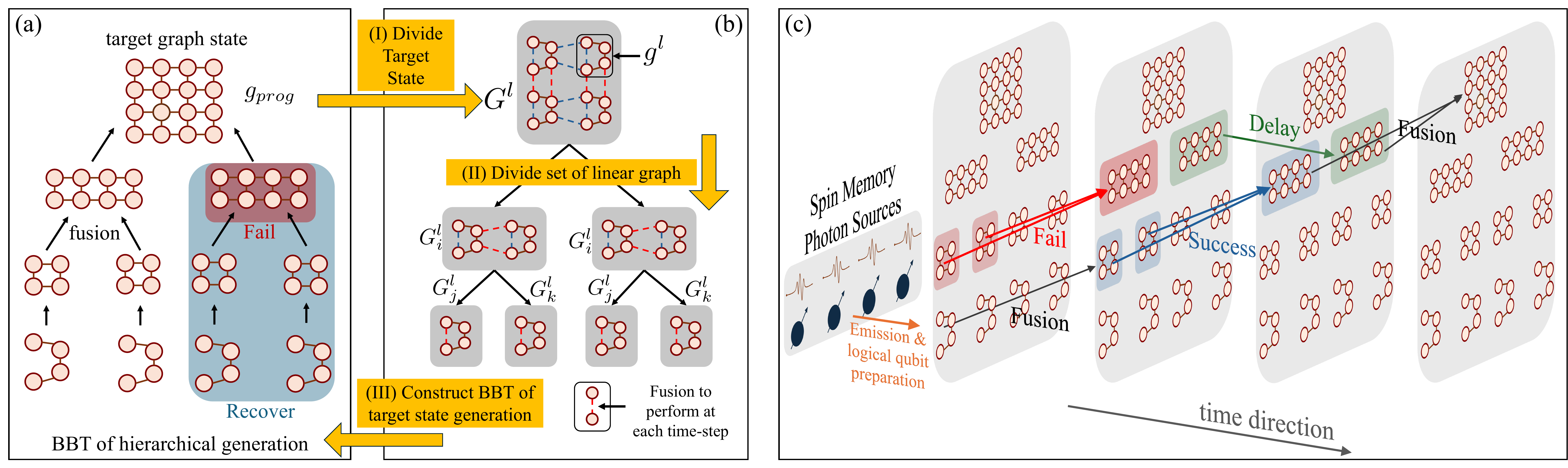}
    \caption{Details of our MemTree compiler.
    (a) The hierarchical generation of target state based on BBT.
    (b) Our compiler framework for building BBT.
    (c) The overall pipeline for target state generation, from a time direction prospective. Each slice corresponds to a time step in the cycles.}
    \label{fig:compiler}
    \vspace{-2mm}
\end{figure*}

\vspace{3mm}
\section{The MemTree Compilation Framework}
\label{sec:compiler}
\subsection{Hierarchical Generation of Target State}
In this section, we introduce our compilation scheme, which generates the target graph state from a set of primitive caterpillar states.
We adapt a hierarchical generation method in which the generation process is modeled in a balanced binary tree (BBT), as shown in Fig.~\ref{fig:compiler}(a).
The process starts from the leaves of BBT, which are linear graphs of logical qubits in our tree-encoding.
Each pair of graph states is combined through fusions, forming a larger graph state as their parent.
All fusions in the same layer are simultaneously operated on in one time step, and these time steps are performed sequentially from the layer of leaves (linear graphs) to the root (target graph state).

Here, we explain the reason for designing this hierarchical generation method.
An alternative straightforward method is to apply all the fusions in one time step and directly generate the target graph from linear subgraphs.
Considering the errors of fusion, this straightforward method is not practical:
For example, even with an extremely high fusion success rate $S_{fusion}$ (assuming $S_{fusion}=0.99$), generating the target state of the 100-qubit VQE program requires $k>1000$ times of fusion, leading to a success rate ${S_{fusion}}^k\sim1e^{-5}$.
On the contrary, in our hierarchical generation method, if any fusion operation -- as one node in the BBT fails, we only need to recover a sub-tree with that node as the root, as shown in Fig.~\ref{fig:compiler} (a).

In our hierarchical generation method, the upper bound of generation overhead depends on a \textit{critical path}.
The \textit{Critical path} is a path from a leaf to the root in BBT, which has the maximal total number of fusions along the path.
The algorithm we introduce in the next subsection aims to reduce the critical path overhead.
Overall, we use the balanced binary tree (BBT) to achieve a tradeoff between the overall success rate and execution time.

\subsection{Building the Generation Tree of Target State}
Here, we describe the details of our algorithm for building the balanced binary tree (BBT) of target state generation (Fig.~\ref{fig:compiler}(b)).
Our algorithm is composed of two parts: 
(1) Dividing the target graph state into linear subgraphs, with a minimal number of total fusion operations (yellow box (I)). 
(2) Building the BBT while reducing the overhead on the critical path as much as possible (yellow box (II)-(III)).

\subsubsection{Dividing Target State} 
We utilize the mix-integer-programming (MIP) solver in Gurobi to solve this problem.
First, we model the program graph state as an undirected graph $g_{prog}$, with each qubit as a vertex $v\in V$, and each CZ-entanglement between qubits as an edge $e(i,j) \in E$.
Then, we model the divided subgraphs as $g^l \subseteq g_{prog}$, and define specific constraints to ensure they are linear graphs.
Next, we set the objective as the number of fusions to combine all the $g^l$ into $g_{prog}$, to find its minimal value.
We list the parameters, the constraints, and the objective function in the MIP model.

In the MIP model, we set each $e_{(v_1,v_2)}$ (CZ-entanglement) as a binary \ul{variable}, with its value indicating whether it is cut or preserved:
$$
x_{e,v_1,v_2}=\left\{
\begin{aligned}
1 & \text{, if } e \text{ preserved in subgraphs}\\
0 & \text{, if } e \text{ cut for fusion.} \\
\end{aligned}
\ , \forall e_{(v_1,v_2)}\in E
\right.
$$
We add the \ul{constraint} to ensure the subgraphs are linear -- each vertex $v$ should have its degree $deg(v) \leq 2$:
\begin{equation} \label{eq:cut_constr}
\sum_{v\in\{v_1,v_2\}} \forall x_{e,v_1,v_2} \leq 2, \quad \forall v \in V
\end{equation}
The model's \ul{objective function} is the total number of edge cuttings:
\begin{equation} \label{eq:cut_obj}
K = |E| - \sum_{e_{(v_1,v_2)} \in E} x_{e,v_1,v_2}
\end{equation}
Overall, the MIP model can be formulated as
$$minimize\ objective\ K\ (Eq.~\ref{eq:cut_obj})$$
$$s.t.\ constraint\ Eq.~\ref{eq:cut_constr}$$
Since the constraint Eq.~\ref{eq:cut_constr} may lead to a cyclic linear graph, we apply post-processing on the subgraphs $g^l$, cutting one of its edges to make it acyclic.
Furthermore, we cut these $g^l$ into smaller linear subgraphs to comply with the maximum length of the caterpillar graph allowed in the specific hardware configuration.
Finally, we obtain a set of linear subgraphs $G^l=\{g^l\}$ that can be resembled into $g_{prog}$.

\subsubsection{Constructing BBT of Target State Generation}
We construct the BBT by growing its layers hierarchically from the root $G^l$. 
Each node of the BBT is a subgraph of $g_{prog}$, and this subgraph is composed of a set of linear graphs $G^l_i$, which are a subset of $G^l$.
During the construction, the subset $G^l_i$ in each node is divided into two smaller subsets $G^l_j$ and $G^l_k$, which are grown as the children of $G^l_i$.

For maximally reducing the overhead of \textit{the critical path}, we adopt a straightforward but effective strategy: 
Starting from the root $G^l$, we search for each division from $G^l_i$ to $G^l_j,G^l_k$ with the minimal number of edges to cut.
This strategy ensures that the division in the lower layers (closer to the root) has fewer fusions, as it is more likely to be involved in \textit{the critical path}.
In the meantime, we retain the balance of BBT to minimize its tree-height: when we divide $G^l_i$ into $G^l_j,G^l_k$, the difference in cardinality between the sets $G^l_j$ and $G^l_k$ should not exceed a certain value.
Based on the above strategies, we define another MIP model for dividing each $G^l_i$.

For each linear subgraph $g^l$ that $g^l\in G^l_i$, we define a \ul{variable} to determine whether it is divided into $G^l_j$ or $G^l_k$:
$$
y_{g^l} = \left\{
\begin{aligned}
1 & \text{, if } g^l \text{ divided to } G^l_j\\
0 & \text{, if } g^l \text{ divided to } G^l_k\\
\end{aligned}
\quad , \forall g^l\in G^l_i
\right.
$$
The following \ul{constraints} are used to restrict the difference in cardinality between $G^l_j$ and $G^l_k$:
\begin{equation} \label{eq:div_constr1}
    |G^l_j|\ \geq\ 2^{\lfloor log_2(|G^l_i|) \rfloor}
\end{equation}
\begin{equation} \label{eq:div_constr2}
    |G^l_k|\ \geq\ 2^{\lfloor log_2(|G^l_i|) \rfloor}
\end{equation}
The \ul{objective function} for this MIP model is the number of edge cuts needed to divide $G^l_i$ into $G^l_j$ and $G^l_k$:
\begin{equation} \label{eq:div_obj}
    L = \sum_{
    v_1 \in g^l_1\ \land \ v_2 \in g^l_2,\ \forall v_1,v_2} 
    |y_{g^l_1} - y_{g^l_2}|, 
    \quad\forall e_{(v_1,v_2)}\in E
\end{equation}
Overall, the MIP model can be formulated as
$$minimize\ objective\ L\ (Eq.~\ref{eq:div_obj})$$
$$s.t.\ constraint\ Eqs.~(\ref{eq:div_constr1},\ref{eq:div_constr2})$$
The node dividing process runs recursively until it reaches the leaf node where $G^l_j=G^l_k=1$.
Finally, we can build our BBT of the generation process: the edges cut from $G^l_i$ to $G^l_j,G^l_k$ in each layer represents the number of fusion operations that need to be performed at each time step.

These two MIP models described above have $O(|E|)$ and $O(|G^l_i|)$ complexities in terms of the number of binary variables, which allows for a relatively low compilation runtime. 
In Sec.~\ref{sec:evaluation} we evaluate the overall runtime of our algorithms.

\subsection{Pipeline for Target State Generation} \label{subsec:pipeline}
With the BBT of generation described above, in Fig.~\ref{fig:compiler}(c) we illustrate the generation process from a time-directional perspective.
Similar to the time-like model in OneAdapt\cite{zhang_oneadapt_2025}, the photon source iteratively generates caterpillar states, forming a pipeline for generating target states.
First, the emitted caterpillar states are prepared into the linear graphs of tree-encoded logical qubits. 
Then, at each time step, each layer of subgraphs performs fusion operations to merge into their parent subgraphs while being forwarded in the pipeline.
When the fusion into a subgraph fails (Fig.~\ref{fig:compiler}(c) red arrows), its sibling subgraph is delayed to the next time step (green arrow) and waits for the next successful generation of this subgraph (blue arrows).
The descendants of this sibling subgraph are also delayed.
Overall, we aim to generate as many target states as possible, thereby maximizing the execution shots of the quantum program within the limited time cycles of the pipeline.

\section{Experimental Methodology}

\label{sec:methodology}
\subsection{Baselines}
\subsubsection{Boosted Fusion Schemes}
In the first part, we select baselines within the same architecture -- quantum spin memory. 
We evaluate our tree-encoded fusion scheme through comparison with two mainstream boosted-fusion schemes introduced in Sec.~\ref{sec:encoding}.A: 
the \textbf{redundantly-encoded fusion}\cite{hilaire_near-deterministic_2023} and \textbf{repeat-until-success (RUS) fusion}\cite{lim_repeat-until-success_2005,gliniasty_spin-optical_2024}. 
We implement their up-to-date protocols according to the most recent research\cite{chan_tailoring_2025} and integrate them into our compiler framework.
Based on the analysis and simulation from the papers\cite{hilaire_near-deterministic_2023,chan_tailoring_2025} of these schemes, we choose the code sizes $m_{Redun}=5$ and $m_{RUS}=6$ for optimal error-tolerance performance.

\subsubsection{SOTA MBQC Compiler}
In the second part, we select the baseline from other MBQC architectures, namely the all-photonic and emitter-based architectures.

\begin{figure}[t]
    \centering
    \includegraphics[width=0.4\textwidth]{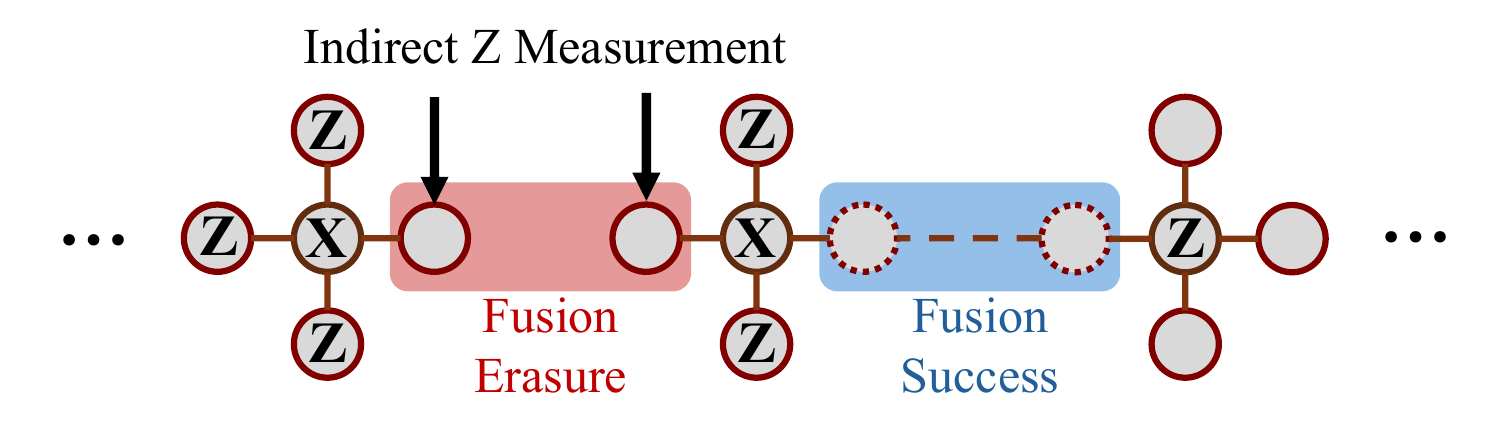}
    \caption{Addressing erasure error in OneAdapt-ET.}
    \vspace{-3mm}
    \label{fig:oneadaptet}
\end{figure}

We compare our framework with the SOTA compiler of the all-photonic architecture -- \textbf{OneAdapt}\cite{zhang_oneadapt_2025}.
Furthermore, we improve OneAdapt by designing an erasure-tolerance scheme and integrating the scheme into it, namely \textbf{OneAdapt-ET}.
OneAdapt-ET addresses the qubit that undergoes fusion erasure by applying an indirect Z-measurement based on the graph state property introduced in Sec.~\ref{sec:encoding}.B.
Specifically, we apply an X-measurement on a neighboring free qubit that is not involved in the normalization path, then apply a Z-measurement on all other adjacent qubits of this neighboring free qubit.
The scheme of OneAdapt-ET is depicted in Fig.~\ref{fig:oneadaptet}.

For a fair comparison, we set the configuration of \textit{the time-like edge length limit} $D_f$ in OneAdapt and OneAdapt-ET as $D_f=30$ virtual layers, strictly following the evaluation settings in \cite{zhang_oneadapt_2025}.
A recent experimental demonstration\cite{psiquantum_team_manufacturable_2025} from PsiQuantum claims a 125 MHz photon source pumping rate in their all-photonic architecture. 
While in OneAdapt, each virtual layer includes 4 physical photon resource layers ($PL=4$), and the maximal duration of the time-like edge is 960 ns.
Correspondingly, we set our maximal delay to 32 emission layers, since the maximal emission time of each caterpillar state layer is 30 ns, according to the experiments in \cite{hilaire_near-deterministic_2023}.
Additionally, we set the resource state layer (RSL) of OneAdapt/OneAdapt-ET as $14n\times14n$ 2D size, according to the source code of OneAdapt.

We select the SOTA compilation framework \textbf{RLGS}\cite{li_reinforcement_2025} for the emitter-based architecture.
Although the emitter-CZ operation is still out of reach in real experiments, we can still compare it as a long-term future architecture.
Due to the distinctive hardware of emitter-based architecture, RLGS uses a set of different metrics\cite{li_reinforcement_2025}.
Hence, we compare with RLGS specifically on fidelity metrics reported in their paper: 
(1) Fidelity affected by decoherence error ($F_{de}$), and (2) Fidelity affected by emitter-CZ ($F_{CZ}$), which corresponds to the fidelity affected by fusion ($F_{fus}$) in our framework.

\subsection{Benchmark Programs}
We select a set of benchmark programs, including the Bernstein–Vazirani algorithm (BV), the Quantum Approximate Optimization Algorithm (QAOA), Grover's Algorithm (Grover), the Quantum Fourier transform (QFT), quantum Hamiltonian simulation (QSIM), the Ripple Carry Adder (RCA), and the Variational Quantum Eigensolver (VQE). 
In the comparison with redundantly-encoded and RUS fusion schemes, we set the size of the benchmark program from 2-qubits to 20-qubits. 
This is because these two baselines of fusion schemes have a prolonged execution time, which is out of reach in simulation.
For the comparison between our compiler and OneAdapt\cite{zhang_oneadapt_2025}, we use exactly the same benchmark programs and settings, with program sizes of 36, 64, and 100-qubits.

\begin{figure*}[h]
    \centering
    \includegraphics[width=0.9\textwidth]{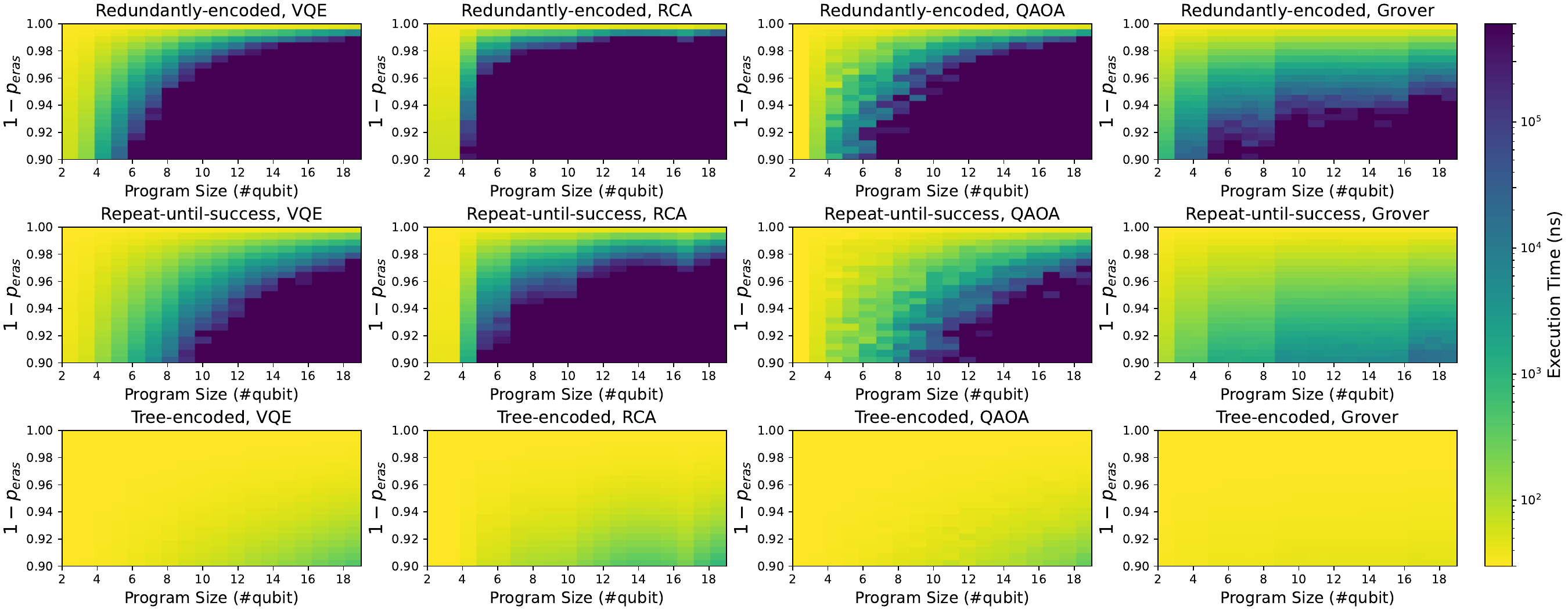}
    \caption{Execution time comparison between tree-encoded scheme and baselines.}
    \label{fig:encoding-compare1}
    \vspace{-3mm}
\end{figure*}

\begin{figure}[h]
    \centering
    \includegraphics[width=0.45\textwidth]{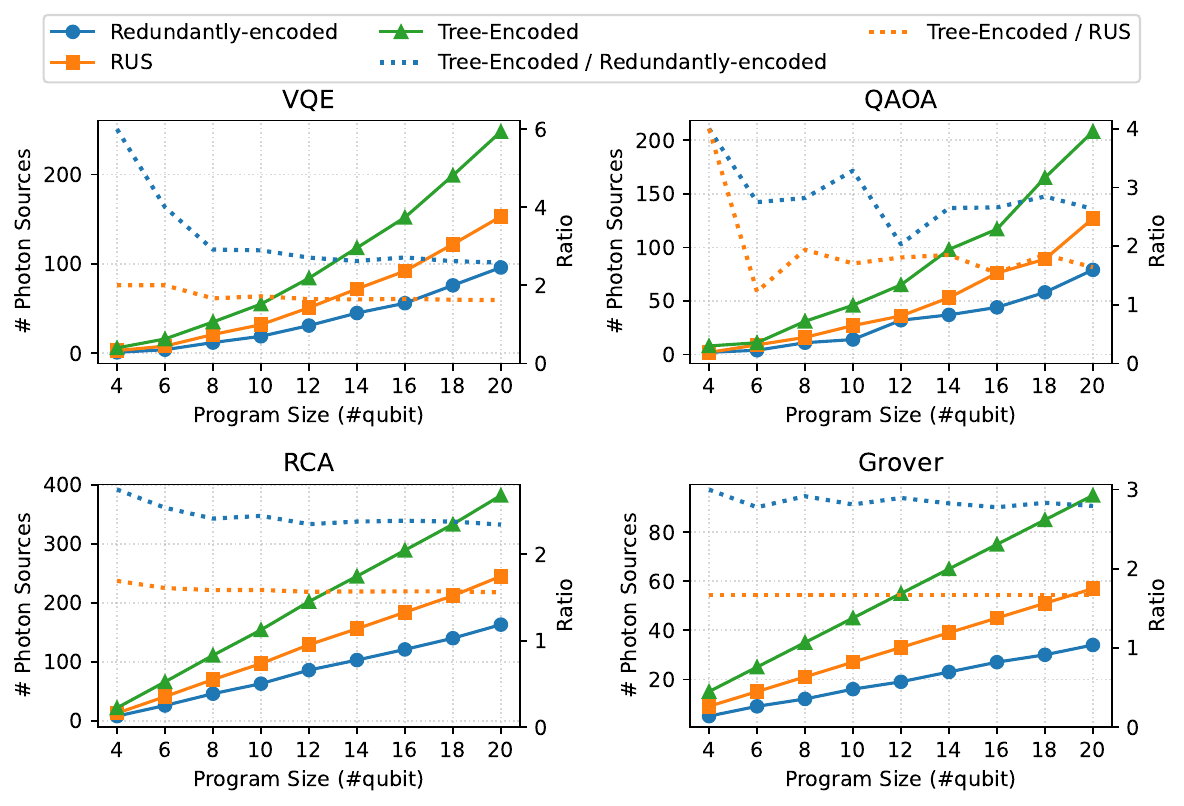}
    \caption{Number of required photon sources comparison between tree-encoded scheme and baselines.}
    \label{fig:encoding-compare2}
    \vspace{-3mm}
\end{figure}

\begin{table}[b]
\vspace{-5mm}
\rev{\caption{Details of noise model we adapt on the fidelity comparison.}\label{tab:noise}}
\centering
\rev{\begin{tabular}{|c|c|c|c|}
\hline
Compiler                & OneAdapt    & RLGS                                  & MemTree     \\ \hline
Based on Platform       & PsiQuantum  & \cite{russo_photonic_2018} (Simulation) & Quandela    \\ \hline
Dephasing $T_2$         & 2.04 $\mu$s & 4.4 $\mu$s                            & 2.34 $\mu$s \\ \hline
CZ (Fusion) Fidelity & 99.75\%      & 99\%                                  & 99\%        \\ \hline
$t_{cycle}$ & 8 ns & 10 ns (emitter-CZ) & 30 ns \\ \hline
\end{tabular}}
\end{table}

\subsection{\rev{Noise Model}}
\subsubsection{\rev{Fusion Failure and Erasure Errors}}
Here are the details of our simulator for spin memory architecture PQC.
Based on recent experimental works on spin memory architecture\cite{maring_versatile_2024,huet2025deterministic} and linear-optical PQC\cite{bartolucci_fusion-based_2023,aghaee_rad_scaling_2025,psiquantum_team_manufacturable_2025}, we simulate the following important errors in PQC: fusion failure and erasure errors, photon source decoherence, and fusion infidelity (indistinguishability).
We set $1-p_{fail}=0.75$ as the fusion success rate when assuming no erasure error, which corresponds to the error model introduced in Sec.~5.1 of the OneAdapt paper\cite{zhang_oneadapt_2025}.
This success rate can be achieved by utilizing additional interferometric setups reported in previous works\cite{grice2011arbitrarily,ewert20143,olivo2018ancilla}.

\subsubsection{\rev{Decoherence Errors}}
We simulate the \ul{photon source (emitter) decoherence} based on $F_{de}=e^{\frac{-N_e T_{gen}}{T_2}}$, in accordance with the error model used in RLGS\cite{li_reinforcement_2025}.
\rev{
We set the dephasing time of RLGS based at $T_2=4.4\mu s$, as reported by \cite{huthmacher2018coherence,li_reinforcement_2025}.
As for OneAdapt and MemTree, we estimate the dephasing time based on the Bell state (GHZ state) fidelity reported in corresponding hardware demonstration\cite{psiquantum_team_manufacturable_2025, huet2025deterministic}.
In \cite{psiquantum_team_manufacturable_2025} the fidelity of a 2-qubit Bell state is $99.22\%$ for all-photonic architecture, while \cite{huet2025deterministic} reports a $95\%$ optimal fidelity of a 4-qubit GHZ state for spin memory architecture.
The dephasing time $T_2$ can be calculated by
$$T_2=\frac{-N_qt_{gen}}{\ln(F_{state})},\ N_q=\text{\#qubit},\ t_{gen}=\text{generation time.}$$
The dephasing time for each architecture is listed in Table.~\ref{tab:noise}.
}
\subsubsection{\rev{Coherent Errors of Fusion Operation}}\label{subsec:coherr}
We simulate the overall \ul{fusion fidelity} $F_{fus}={\sigma_{fus}}^{N_{fus}}$, corresponding with $F_{CZ}={\sigma_{CZ}}^{N_{CZ}}$ reported in RLGS\cite{li_reinforcement_2025}.
\rev{
We set the fidelity of each emitter-CZ operation at $\sigma_{CZ}=99\%$ for RLGS,
as reported by \cite{russo_photonic_2018} in the form of pulse-level simulation result.
Based on the \textit{Hong-Ou-Mandel (HOM)} visibility $V_{HOM}=99.5\%$ reported in \cite{psiquantum_team_manufacturable_2025}, we set the fidelity of type-II fusion operation at $\sigma_{fus}=\frac{1+V_{HOM}}{2}=99.75\%$ based on \cite{hong1987measurement}.
Meanwhile, we set the OSRP fidelity for spin memory at 99\%, as reported in \cite{huet2025deterministic}.
}

\subsection{\rev{MemTree} Simulator Configurations} \label{subsec:simconf}
We simulate the generation of caterpillar states according to hardware configurations reported in \cite{maring_versatile_2024,huet2025deterministic}.
Specifically, each qubit in a caterpillar state is emitted through an excitation pulse of InGaAs semiconductor quantum-dots, while assisted by an optical spin rotation pulse (OSRP) to define the caterpillar structure\cite{huet2025deterministic}.
Generation of a caterpillar graph state includes a 12 ns initialization time, plus a 0.6 ns time cycle for the emission of each qubit.
The near-term spin memory technique can produce a caterpillar state with at most 30-qubit\cite{huet2025deterministic}, which is set as the \ul{maximal size of the caterpillar} in our framework.
For calculating the \ul{average execution time}, we simulate $2\times10^4$ cycles of caterpillar state emissions and divide the total time by the number of successful shots executed during these cycles.
In addition, we choose $b=4$ as the tree-encoding parameter, based on the parametric study in Sec~\ref{sec:evaluation}.D.

\subsection{Metrics}
We evaluate the performance of our compiler using the following metrics: average execution time of quantum programs, number of photon sources, compilation runtime, and fidelity of the quantum program.
\rev{For fidelity, we include decoherence fidelity $F_{de}$, and CZ (fusion) fidelity $F_{CZ}~(F_{fus})$.}

\begin{figure*}[t]
    \centering
    \includegraphics[width=0.95\textwidth]{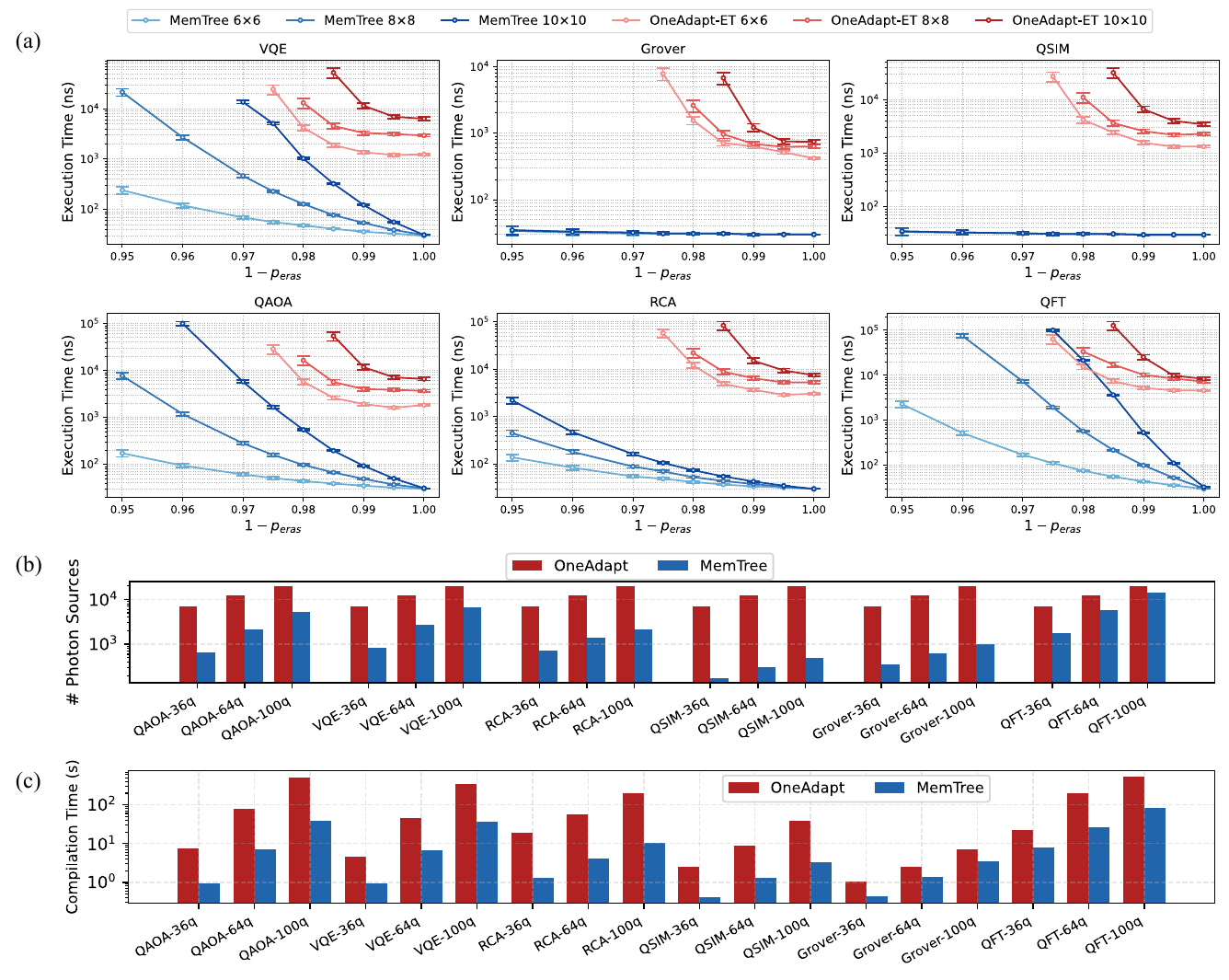}
    \caption{Comparison of MemTree with OneAdapt\cite{zhang_oneadapt_2025} and OneAdapt-ET. (a) The average execution time of quantum programs, when $p_{eras}=0$, the results are evaluated on OneAdapt without erasure-tolerance strategy. \rev{The error bars represent the value range with a statistical 95\% CI (confidence interval), over 1000 times of experiment and each with $2\times10^4$ shots}. (b) Number of required photon sources. (c) Total compilation runtime of compilers.}
    \label{fig:compiler-compare1}
    \vspace{-5mm}
\end{figure*}

\begin{figure}[h]
    \centering
    {\includegraphics[width=0.45\textwidth]{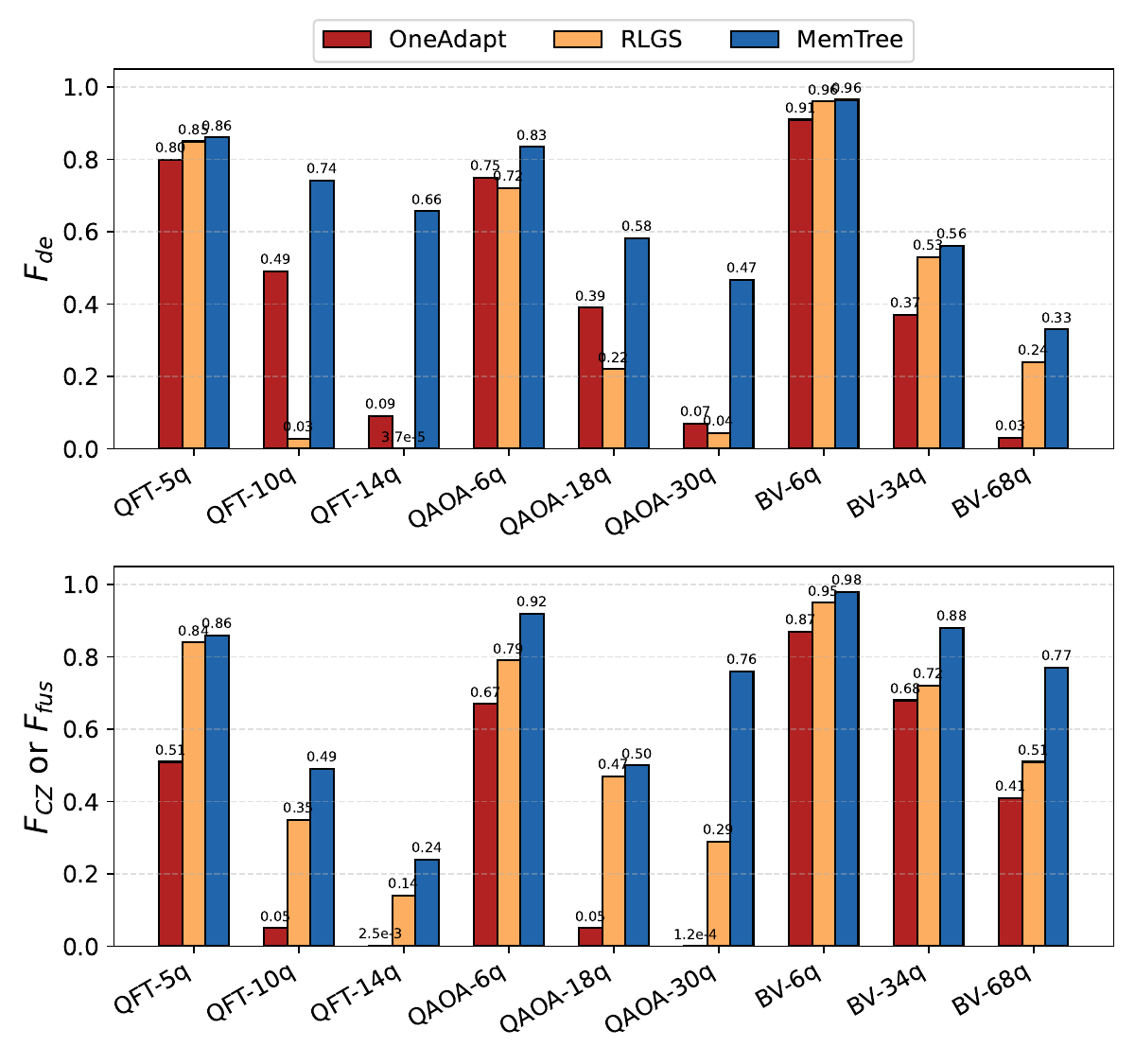}}
    \caption{Comparison \rev{on decoherence errors and CZ errors} between \rev{OneAdapt\cite{zhang_oneadapt_2025},} RLGS\cite{li_reinforcement_2025} and MemTree.}
    \label{fig:compiler-compare2}
    \vspace{-3mm}
\end{figure}

\section{Evaluation}
\label{sec:evaluation}
\subsection{Comparison with Boosted-Fusion Schemes}
Fig.~\ref{fig:encoding-compare1} and Fig.~\ref{fig:encoding-compare2} present the comparison of our tree-encoded fusion scheme with the redundantly-encoded and RUS fusion schemes under the hardware configurations of the quantum spin memory architecture.
In this comparison, all fusion schemes are integrated in MemTree with the same compilation algorithm.
While fixing the fusion failure rate $p_{fail}=0.25$ (thus $1-p_{fail}=0.75$), we compare the program execution time and the number of required photon sources.
The program size (\#qubit) varies from 2-qubit to 20-qubit, and the erasure rate during fusion ($p_{eras}$) varies from 0\% to 10\%.
Due to the extremely large simulation overhead when the program size scales up, we truncate the execution time to at most $6\times10^5$ ns. 
Fig.~\ref{fig:encoding-compare1} shows that our scheme significantly reduces the average execution time of quantum programs, gaining an average reduction rate of $1.9\times10^{-3}$ and $1.7\times10^{-2}$, compared to redundantly-encoded and RUS, respectively.
Fig.~\ref{fig:encoding-compare2} shows that our scheme consumes more photon sources than the baseline schemes, with an average of $2.55\times$ and $1.63\times$ compared to redundantly-encoded and RUS, respectively.
Nevertheless, considering the exponential reduction in execution time, we believe that the tree-encoded scheme is an appropriate strategy for \textit{trading space for time}.
Besides, it can be observed that for tree-encoded fusion, its disadvantage on photon sources decreases as the \#qubit grows (Fig.~\ref{fig:encoding-compare2} dotted lines).

\subsection{Comparison with SOTA Compilers of Other Architectures}
\textbf{\rev{Execution Time.}} 
Fig.~\ref{fig:compiler-compare1}(a)-(c) present the comparison of our framework MemTree with OneAdapt and OneAdapt-ET on average execution time, the number of photon sources, and compilation runtime.
In Fig.~\ref{fig:compiler-compare1}(a), the execution time results on benchmarks with 36, 64, and 100-qubits are shown, with varying fusion erasure rates $p_{eras}$ from 0\% to 5\%.
Note that the realistic $p_{eras}$ estimated from the hardware experiment is on the order of $\sim1\%$\cite{psiquantum_team_manufacturable_2025}.
We set a simulation limit for the execution time ($2\times10^5$ ns), since a longer execution time requires $>80$ hours of simulation on our machine.
The results show that our compiler framework achieves an exponential improvement in execution time, and the reduction rates are $1.5\times10^{-2}$, $1.1\times10^{-2}$, $3.8\times10^{-2}$, $5.6\times10^{-3}$, $1.1\times10^{-2}$, and $8.8\times10^{-3}$ for VQE, QAOA, Grover, RCA, QSIM, and QFT, respectively.
Shown in Fig.~\ref{fig:compiler-compare1}(b)-(c), the number of photon sources is reduced to $0.18\times$ on average, and the compilation time is reduced $0.14\times$ on average compared to OneAdapt.
For the Grover and QSIM benchmarks, the intrinsic structure of their graph state leads to a relatively low number of fusions when divided into caterpillar states, and this number does not scale with program size, while larger programs only require more photon sources.

\textbf{\rev{Circuit Fidelity.}}
Fig.~\ref{fig:compiler-compare2} presents the comparison of MemTree with RLGS\cite{li_reinforcement_2025} on $F_{de}$ and $F_{CZ}/F_{fus}$, using the benchmark results (QFT, QAOA, BV) reported in their paper.
Note that the fusion operation in our architecture behaves similarly to emitter-CZ in the emitter-based architecture of RLGS, thus we compare our $F_{fus}$ with their $F_{CZ}$.
The results show that we achieved a significant improvement on $F_{de}$, especially an exponential enhancement for QFT and QAOA. 
\rev{The results show that MemTree outperforms OneAdapt and RLGS both in $F_{de}$ and $F_{CZ}$, and the advantage grows with \#qubit.}

\subsection{Ablation Study}\label{subsec:ablation}
\begin{figure}[t]
    \centering
    \includegraphics[width=0.48\textwidth]{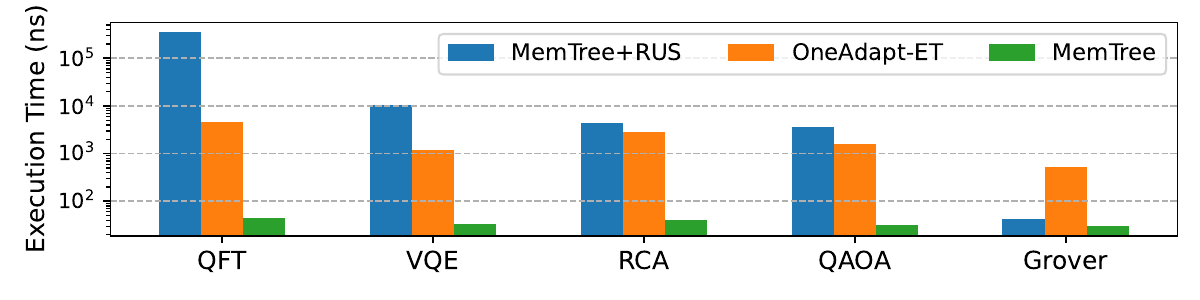}
    \vspace{-2mm}
    \caption{Encoding parameter study and ablation study.}
    \label{fig:study}
\end{figure}

We conduct the following ablation study to verify that our performance gains over OneAdapt primarily stem from our novel tree-encoded fusion rather than merely from differences in PQC architecture or hardware configurations.
Here, we compare three different compiler settings: MemTree with the RUS fusion scheme (the same we used in Sec.~\ref{sec:evaluation}.A), MemTree, and OneAdapt-ET, on 36-qubit benchmarks with $p_{eras} = 0.5\%$.
The results are shown in Fig.~\ref{fig:study}(b).
It can be observed that when the Tree-encoded fusion is replaced by the RUS fusion scheme, it under-performs OneAdapt on all benchmarks except Grover.
This ablation experiment supports for the novelty and effectiveness of our design of tree-encoded fusion scheme.

\rev{
\subsection{Trade-Off Analysis of System Characterizations}
In Table.~\ref{tab:freq_noise} we analyze the system characterization of each compiler with the following metrics: \textit{emission frequency} = \#(physical photon) /ns, \textit{CZ frequency} = \#(CZ or fusion operation) /ns, and \textit{utilization rate} = \#(logical qubit) / \#(total physical photon).
From the data we can find MemTree establishing a proper trade-off between operation frequency and photon utilization rate.
(i) Higher frequency (OneAdapt) leads to shorter QPU runtime, but requires larger number of noisy operations, hence the dominant errors derive from fusion operations.
(ii) Higher utilization rate (RLGS) is based on fewer and more reliable CZ operations, but leads to lower frequency and longer QPU runtime, hence the dominant errors is decoherence.
(iii) Based on the spin memory architecture and tree-encoded scheme, MemTree is designed to reach a trade-off between frequency and logical CZ (fusion) reliability. 
This prevents extremely high error rate deriving from CZ (fusion) or decoherence, thus MemTree outperforms both baselines in the evaluation of Fig.\ref{fig:compiler-compare2}.
}
\begin{table}[h]
\vspace{-3mm}
\centering
\rev{
\caption{Frequency-Noise Analysis (on 30-qubit QAOA)} \label{tab:freq_noise}
\begin{tabular}{|c|c|c|c|}
\hline
Compiler         & OneAdapt & MemTree         & RLGS        \\ \hline
Emission Frequency (/ns) & $\approx 2 \times 10^3$ & $\approx 7 \times 10^2$ & $\approx 10$ \\ \hline
CZ Frequency (/ns)       & $\approx 1 \times 10^3$ & $\approx 2 \times 10^2$ & $\approx 1$  \\ \hline
Utilization rate & $\approx 0.03\%$    & $\approx 10\%$           & 100\%       \\ \hline
Dominant Error   & Fusion   & F-D Tradeoff & Decoherence \\ \hline
\end{tabular}
}
\end{table}

\vspace{-2mm}
\subsection{Feed-Forward Control in PQC System}
There are two cases in MemTree where feed-forward control is required.
(i) In the tree-encoded fusion scheme, the measurement basis of ancillary qubits is updated according to the fusion outcome to handle fusion failure and erasure.
This feed-forward is not on the critical optical path: once a fusion outcome is detected, the affected qubits can remain as dangling qubits, and the controller only needs to record which recovery pattern will later be applied.
Therefore, the corrective measurements do not need to be triggered immediately after fusion; they only need to be synchronized before the dangling branch is consumed by later graph-state measurements, or before the final measurement stage of the quantum program, following the adaptive-measurement model of MBQC~\cite{varnava_loss_2006}.

(ii) In the graph-generation pipeline (Sec.~\ref{subsec:pipeline}), feed-forward is also needed to decide whether a sub-graph should be delayed to the next timestep after an unsuccessful fusion.
This control path consists of photon detection, a small classical decision circuit, and the timing/delay module.
In our target hardware, the measurement signal is produced by \textit{superconducting nanowire single-photon detectors} with latency below $50$ ps~\cite{marsili2013detecting}; the detector outputs are then passed to a small combinational logic block (e.g., a $b$-input AND/OR network for the $b$ fusion branches), which decides whether the logical fusion succeeds or whether the sibling sub-graph must be stalled.
The resulting control signal drives the time-delay module.
We estimate the total classical feed-forward latency to be below $5$ ns, and implement this logic using \texttt{FFCircuitProvider} in Perceval~\cite{heurtel2023perceval}.
Since this latency is well below one emission timestep in spin-memory hardware, the updated measurement pattern and sub-graph schedule can be synchronized before the next emission layer begins.

\begin{figure}[b]
    \centering
    \includegraphics[width=0.4\textwidth]{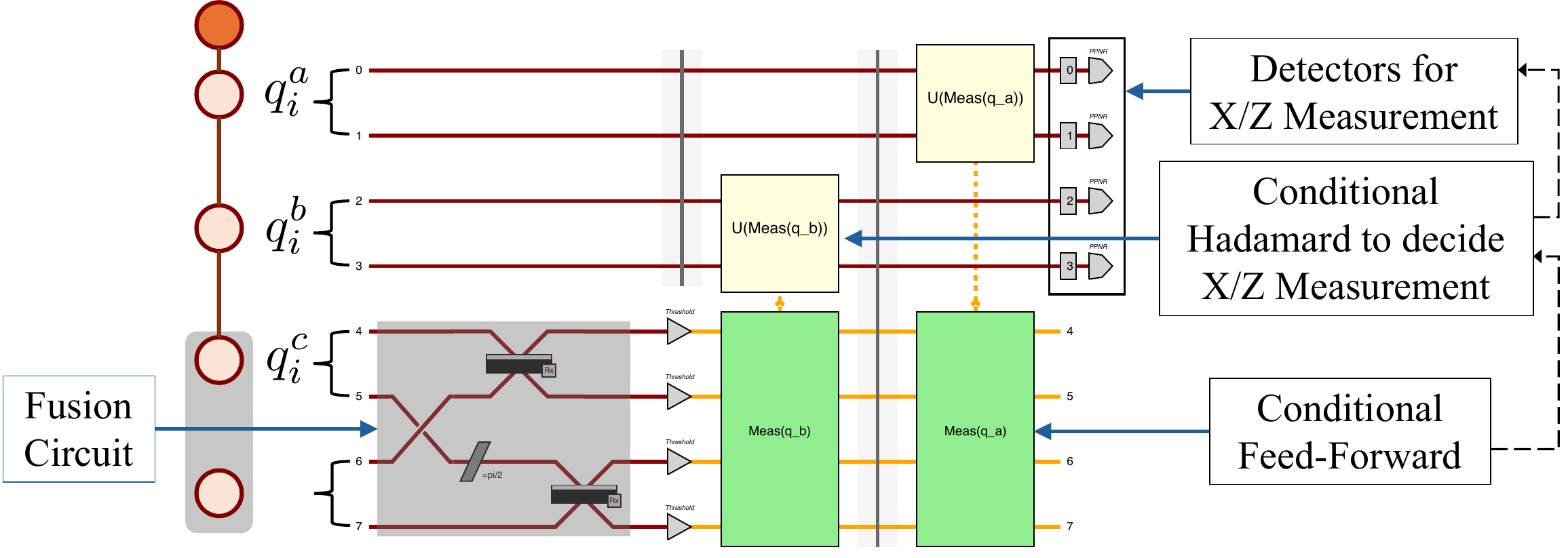}
    \caption{The optical hardware circuit for tree-encoded fusion.}
    \label{fig:realhw}
    \vspace{-5mm}
\end{figure}

\subsection{Real Photonic Hardware Experiment} \label{subsec:realexp}
\rev{Here we perform a small-scale experiment on real photonic quantum hardware~\cite{quandela_cloud}.}
In this experiment, the optical hardware circuit is built with the Perceval PQC toolkit\cite{heurtel2023perceval}.
We illustrate the most important part of the hardware circuit in Fig.~\ref{fig:realhw}, which is the fusion operation and dealing with possible fusion failure or erasure.
In the circuit, each qubit is represented by a dual-rail encoding -- two photon modes (e.g., H or V polarization) are used to encode one qubit.
As in Fig.~\ref{fig:realhw}, the fusion circuit is a permutation of photon modes from the two qubits, followed by a phase shift and two beam splitters.
Corresponding to the tree-encoded scheme in Fig.~\ref{fig:treecode}(b), the fusion outcome from $q^c_i$ is detected and triggers a conditional feed-forward operation on $q^a_i$ and $q^b_i$.
The feed-forward operation decides whether to apply an X or Z measurement on $q^a_i$ and $q^b_i$, complying with the error-tolerant measurement patterns.
\rev{The characterization of photonic hardware are as follows: HOM indistinguishability $=92.0\%$, transmittance $=5.16\%$, g2 $=2.0\%$.}

\begin{figure}[t]
    \centering
    \includegraphics[width=0.48\textwidth]{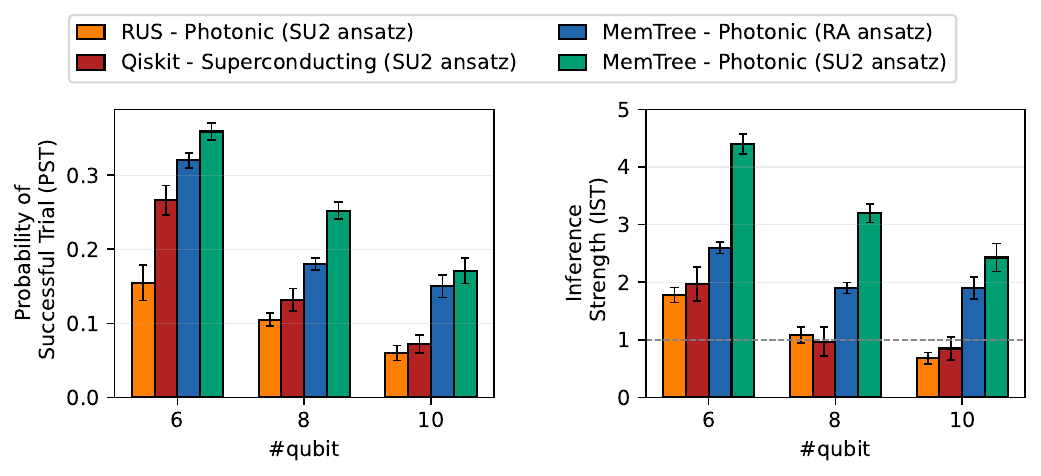}
    \caption{Comparing the performance of QAOA programs on real hardware between superconducting qubits and photonic spin memory. The error bars stand for the standard error (SE).}
    \label{fig:real_qaoa}
    \vspace{-3mm}
\end{figure}

We compile QAOA programs (6--12 qubits) using MemTree, and execute them on photonic hardware. In Fig.~\ref{fig:real_qaoa}, MemTree are compared with repeat-until-success (RUS) scheme\cite{gliniasty_spin-optical_2024} executed on photonic hardware, and Qiskit transpilation\cite{qiskit2024} executed on IBM Torino superconducting quantum computer. 
We use the \textit{EfficientSU2} (SU2) ansatz for QAOA as the default settings, and add a setting of \textit{RealAmplitudes} (RA) ansatz to MemTree to extend the comparison.
Note that in RA ansatz the parameterized rotation gates are restricted to $\text{R}_Y(\theta)$ only, as a simplified ansatz.
The results are evaluated in two metrics, which are \textit{Probability of Successful Trial (PST)}\cite{das2019case,murali2019noise,liu2020reliability,tannu2018case,tannu2019ensemble} and \textit{Inference Strength (IST)}\cite{liu2020reliability,patel2020veritas,tannu2019ensemble}.
From the evaluation results in Fig.~\ref{fig:real_qaoa}, on average, MemTree (SU2 ansatz) achieves an improvement on Probability of Success Trial (PST) by the ratio of 2.68$\times$ compared to [RUS + photonic] and 2.20$\times$ compared to [Qiskit + superconducting]. Also, MemTree achieves an improvement on Inference Strength (IST) by the ratio of 3.23$\times$ compared to [RUS + photonic] and 2.91$\times$ compared to [Qiskit + superconducting].

\rev{From above results, we analyze the reason that PQC hardware outperforms superconducting QPU: PQC has significantly lower crosstalk than matter-based systems, while spin-memory single photon sources are isolated and have no interaction with each others. 
Consequently, PQC provides higher parallelism of CZ operations, and reduce circuit execution time. 
As we protect the fusion (CZ) with tree-encode scheme, the overall quantum noise is efficiently suppressed.}
Generally, SU2 ansatz performs better than RA ansatz for MemTree, however RA starts to outperform SU2 when the number of qubits scales up.
This attributes to the fewer number of parameters in RA ansatz, which leads to lower complexity of optimization when \#qubit scales up.

\section{Related Works and Discussions}
Recent research on quantum computer systems has primarily addressed error correction for superconducting platforms \cite{vittal2023astrea, das2022afs, tannu2019not, ayanzadeh2023frozenqubits} via compilation advances\cite{shi2019optimized, ding2020systematic} and other architectural improvements\cite{maurya2022compaqt, xu2023systems, tomesh2022supermarq, stein2025hetec}. For photonic quantum computing, compilation frameworks target measurement-based systems\cite{zhang2023oneq}, probabilistic fusion operations\cite{zhang2024oneperc}, and bosonic encodings\cite{zhou2024bosehedral}. FCM\cite{mo2024fcm} uses wire cutting to partition circuits and reduce fusion counts through classical post-processing, while our work addresses fusion erasure errors through tree-encoded fusion schemes in spin memory architecture.
\rev{FMCC\cite{li2024fmcc} reduces photonic MBQC cluster-state depth by exploiting flexible mapping variants with dynamic programming and heuristics.}

Prior work on biased-noise QEC, such as the XZZX surface code~\cite{bonilla_ataides_xzzx_2021} and superconducting dual-rail cavity codes~\cite{teoh_dualrail_2023}, studies circuit-model protection by tailoring syndrome-based correction to a hardware-specific error hierarchy. 
By contrast, our setting is fusion-based MBQC on optical photonic graph states, where the dominant challenge arises from imperfect graph-state generation itself. 
In particular, fusion \textit{failure} and fusion \textit{erasure} describe two distinct error modes of the fusion primitive, but they do not form a biased-noise model in the usual QEC sense, since neither is simply a dominant variant of the other. 
Our method corrects graph-generation uncertainty through graph-state measurement patterns and indirect measurements on ancillary qubits, rather than through circuit-level decoding of a biased code.

Beyond spin-memory hardware, the same loss-tolerant logical-fusion idea can also be applied to other PQC architectures whenever graph states are built through fusion. In particular, all-photonic schemes such as fusion-based quantum computation and OneAdapt already rely on small resource states and repeated fusion measurements~\cite{bartolucci_fusionbased_2023,zhang_oneadapt_2025}. In that case, our method can be adapted by replacing the original fusion units with tree-encoded logical qubits, while changing only the resource-state preparation procedure: spin-memory hardware prepares them efficiently from caterpillar states, whereas all-photonic systems would synthesize them from Bell pairs or other small photonic resource states before entering the fusion pipeline. The loss-tolerant recovery mechanism itself remains unchanged, since it still follows graph-state measurement rules and indirect measurements~\cite{varnava_loss_2006}.

\section{Conclusion}
In this work, we present MemTree based on spin-memory architecture, while introducing the tree-encoded fusion to address fusion erasure.
It leads to substantial reduction on execution time and improvement on fidelity, while maintaining a reasonable photon-resource overhead, outperforming SOTA PQC compilers.
Moreover, the experiment on real hardware unleash the potential advantages of PQC, when erasure errors are properly addressed.

\clearpage
\bibliographystyle{IEEEtranS}
\bibliography{ref}

\end{document}